\documentclass[journal]{IEEEtran}
\ifCLASSINFOpdf
\else
\fi
\usepackage{multirow}
\usepackage{array}
\usepackage{booktabs}
\usepackage{subcaption}
\usepackage{colortbl}
\usepackage{amsmath}
\usepackage{soul}
\usepackage{cite}
\usepackage{pifont}
\usepackage{amssymb}
\newcommand{\cmark}{\ding{51}}
\newcommand{\xmark}{\ding{55}}
\usepackage{graphicx}
\usepackage[utf8]{inputenc}
\usepackage[normalem]{ulem}
\usepackage{comment}
\usepackage{balance}
\usepackage{siunitx}
\usepackage{hyperref}

\begin{document}
%
\title{COVID-19 Infection Map Generation and Detection from Chest X-Ray Images}
%
%
%

\author{Aysen Degerli,
        Mete Ahishali,
        Mehmet Yamac,
        Serkan Kiranyaz,
        Muhammad E. H. Chowdhury,
        Khalid Hameed,
        Tahir Hamid, 
        Rashid Mazhar,
        and~Moncef~Gabbouj
\thanks{Aysen Degerli, Mete Ahishali, Mehmet Yamac, and Moncef Gabbouj are with the Faculty of Information Technology and Communication Sciences, Tampere University, Tampere, Finland (e-mail: name.surname@tuni.fi).}
\thanks{Serkan Kiranyaz and Muhammad E. H. Chowdhury are with the Department of Electrical Engineering, Qatar University, Doha, Qatar (e-mail: mkiranyaz@qu.edu.qa and mchowdhury@qu.edu.qa).}
\thanks{Khalid  Hameed  is  an  MD  in  Reem  Medical  Center,  Doha,  Qatar  (e-mail: dr.khalid@reemmedicalcenter.com). }
\thanks{Tahir Hamid is a consultant cardiologist in Hamad Medical Corporation Hospital and with Weill Cornell Medicine - Qatar, Doha, Qatar. Rashid Mazhar is an MD in Hamad Medical Corporation Hospital, Doha, Qatar.}}

\maketitle

\begin{abstract}
Computer-aided diagnosis has become a necessity for accurate and immediate coronavirus disease 2019 (COVID-19) detection to aid treatment and prevent the spread of the virus. Numerous studies have proposed to use Deep Learning techniques for COVID-19 diagnosis. However, they have used very limited chest X-ray (CXR) image repositories for evaluation with a small number, a few hundreds, of COVID-19 samples. Moreover, these methods can neither localize nor grade the severity of COVID-19 infection. For this purpose, recent studies proposed to explore the activation maps of deep networks. However, they remain inaccurate for localizing the actual infestation making them unreliable for clinical use. This study proposes a novel method for the joint localization, severity grading, and detection of COVID-19 from CXR images by generating the so-called \textit{infection maps}. To accomplish this, we have compiled the largest dataset with 119,316 CXR images including 2951 COVID-19 samples, where the annotation of the ground-truth segmentation masks is performed on CXRs by a novel collaborative human-machine approach. Furthermore, we publicly release the first CXR dataset with the ground-truth segmentation masks of the COVID-19 infected regions. A detailed set of experiments show that state-of-the-art segmentation networks can learn to localize COVID-19 infection with an F1-score of 83.20\%, which is significantly superior to the activation maps created by the previous methods. Finally, the proposed approach achieved a COVID-19 detection performance with 94.96\% sensitivity and 99.88\% specificity.
\end{abstract}


%
\IEEEpeerreviewmaketitle

\begin{figure}[t!]
    \centering
    \includegraphics[width=0.46\textwidth]{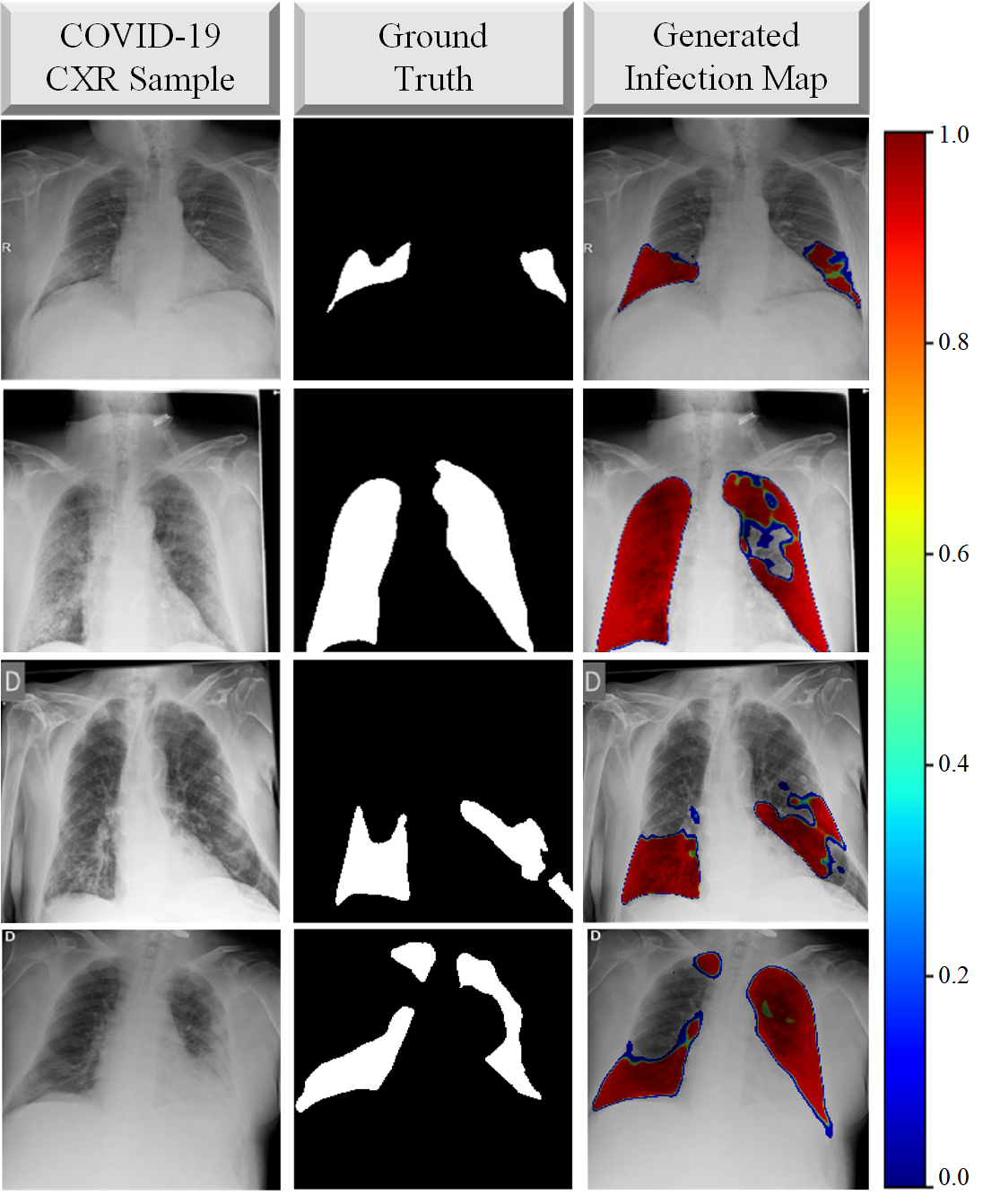}
    \caption{The COVID-19 sample CXR images, their corresponding ground-truth segmentation masks which are annotated by the collaborative human-machine approach, and the generated infection maps from the state-of-the-art segmentation models.}
    \label{fig:GT}
\end{figure}

\section{Introduction}
\IEEEPARstart{C}{oronavirus} disease 2019 (COVID-19) caused by severe acute respiratory syndrome Coronavirus-2 (SARs-CoV-2) was first reported in December 2019 in Wuhan, China. The highly infectious disease rapidly spread around the World with millions of positive cases. As a result, COVID-19 was declared as a pandemic by the World Health Organization in March 2020. The disease may lead to hospitalization, intubation, intensive care, and even death, especially for the elderly \cite{2020coronavirus, world2020coronavirus}. Naturally, reliable detection of the disease has the utmost importance. However, the diagnosis of COVID-19 is not straight-forward since its symptoms, such as cough, fever, breathlessness, and diarrhea are generally indistinguishable from other viral infections \cite{sohrabi2020world, singhal2020review}. 

The diagnostic tools to detect COVID-19 are currently reverse transcription of polymerase chain reaction (RT-PCR) assays and chest imaging techniques, such as Computed Tomography (CT) and X-ray imaging. Primarily, RT-PCR has become the gold standard in the diagnosis of COVID-19 \cite{kakodkar2020comprehensive, li2020stability}. However, RT-PCR arrays have a high false alarm rate which may be caused by the virus mutations in the SARS-CoV-2 genome, sample contamination, or damage to the sample acquired from the patient \cite{tahamtan2020real, xia2020evaluation}. In fact, it is shown in hospitalized patients that RT-PCR sensitivity is low and the test results are highly unstable \cite{li2020stability, xiao2020false, yang2020laboratory, world2020laboratory}. Therefore, it is recommended to perform chest CT imaging initially on the suspected COVID-19 cases \cite{salehi2020coronavirus}, since it is a more reliable clinical tool in the diagnosis with higher sensitivity compared to RT-PCR. Hence, several studies \cite{salehi2020coronavirus, fang2020sensitivity, ai2020correlation} suggest performing CT on the negative RT-PCR findings of the suspected cases. However, there are several limitations of CT scans. Their sensitivity is limited in the early COVID-19 phase groups \cite{bernheim2020chest}, and they are limited to recognize only specific viruses \cite{li2020coronavirus}, slow in image acquisition, and costly. On the other hand, X-ray imaging is faster, cheaper, and less harmful to the body in terms of radiation exposure compared to CT \cite{narin2020automatic, brenner2007computed}. Moreover, unlike CT devices, X-ray devices are easily accessible; hence, reducing the risk of COVID-19 contamination during the imaging process \cite{rubin2020role}. Currently, chest X-ray (CXR) imaging is widely used as an assistive tool in COVID-19 prognosis, and it is reported to have a potential diagnosis capability in recent studies \cite{shi2020review}.

In order to automate COVID-19 detection/recognition from CXR images, many studies \cite{chowdhury2020pdcovidnet, pham2020classification, narin2020automatic, chowdhury2020can, apostolopoulos2020covid, hall2020finding, wang2020covid,  sethy2020detection, zhang2020covid, afshar2020covid} have proposed to use deep Convolutional Neural Networks (CNNs). However, the main limitation of these studies is that the data is scarce for the target COVID-19 class. Such a limited amount of data degrades the learning performance of the deep networks. Two recent studies \cite{yamac2020convolutional} and \cite{ahishali2020comparative} have addressed this drawback with a compact network structure and achieved the state-of-the-art detection performance over the benchmark QaTa-COV19 (initial version) and Early-QaTa-COV19 datasets that consist of $462$ and $175$ COVID-19 CXR images, respectively. Although these datasets were the largest available at that time, such a limited number of COVID-19 samples raises robustness and reliability issues for the proposed methods in general. 

Moreover, all these previous machine learning solutions with X-ray imaging remain limited to only COVID-19 detection. However, as stated by Shi \cite{shi2020large}, COVID-19 pneumonia screening is important for evaluating the status of the patient and treatment. Therefore, along with the detection, COVID-19 related infection localization is another crucial problem. Hence, several studies \cite{yeh2020cascaded, oh2020deep, ozturk2020automated} produced activation maps that are generated from different Deep Learning (DL) models trained for COVID-19 detection (classification) task to localize COVID-19 infection in the lungs. Infection localization has two vital objectives: an accurate assessment of the infection location and the severity of the disease. However, the results of previous studies show that the activation maps generated inherently from the underlying DL network may fail to accomplish both objectives, that is, irrelevant locations with biased severity grading appeared in many cases. To overcome these problems, two studies \cite{alom2020covid_mtnet, haghanifar2020covid} proposed to perform lung segmentation as the first step in their approaches. This way, they have narrowed the region of interest down to the regions of lungs to increase the reliability of their methods. Overall, until this study, screening COVID-19 infection from such activation maps produced by classification networks was the only option for the localization due to the absence of ground-truth of the datasets available in the literature. Many studies \cite{alom2020covid_mtnet, shi2020large, shan2020lung, zhang2020clinically, qiu2020miniseg} have COVID-19 infection ground-truths for CT images; however, ground-truth segmentation masks for CXR images are non-existent.

In this study, in order to overcome the aforementioned limitations and drawbacks, first, the benchmark dataset QaTa-COV19 proposed by the researchers of Qatar University and Tampere University in \cite{yamac2020convolutional} and \cite{ahishali2020comparative} is extended to include $2951$ COVID-19 samples. This new dataset is $3$-$20$ times larger than those used in earlier studies. The extended benchmark dataset, QaTa-COV19 with around $120$K CXR images, is not only the largest ever composed dataset, but it is the first dataset that has the ground-truth segmentation masks for COVID-19 infection regions, as some samples are shown in Fig. \ref{fig:GT}. A crucial property of QaTa-COV19 dataset is that it contains CXRs with other (non-COVID-19) infections and anomalies such as pneumonia and pulmonary edema, both of which exhibit high visual similarity to COVID-19 infection in the lungs. Therefore, this is significantly more challenging task than distinguishing COVID-19 from the normal (healthy) cases as almost all studies in the literature did. 

To obtain the ground-truth segmentation masks for the COVID-19 infected regions, a \textit{human-machine collaborative} approach is introduced. The objective is to significantly reduce the human labor and thus to speed up and also to improve the segmentation masks because when they are drawn solely by medical doctors (MDs), human error due to limited perception, hand-crafting, and subjectivity will deteriorate the overall quality. This is an iterative process, where MDs initiate the segmentation by "manually-drawn" segmentation masks for a subset of CXR images. Then, the trained segmentation networks over this subset generate their own "competing" masks and the MDs are asked to compare them pair-wise (initial manual segmentation vs. machine-segmented masks) for each patient. Such a verification improves the quality of the generated masks as well as the (following) training runs. Over the best masks selected by experts, the networks are trained again this time over a larger set (or even perhaps over the entire dataset), and among the masks generated by the networks, the best masks are selected by the MDs. This human-machine collaboration process continues until the MDs are fully satisfied, i.e., a satisfactory mask can be found among the masks generated by the networks for all CXR images in the dataset. In this study, we show that even with two stages (iterations), highly superior infection maps can be obtained using which an elegant COVID-19 detection performance can be achieved. 

\begin{figure*}[ht!]
    \centering
    \includegraphics[width=1\textwidth]{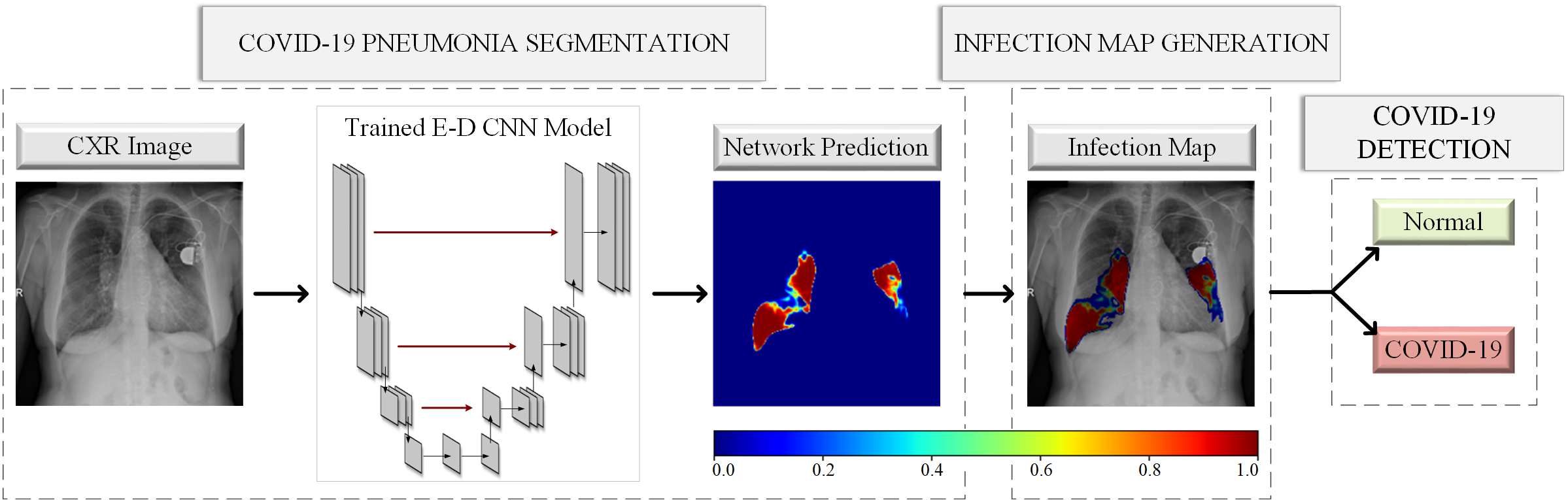}
    \caption{The pipeline of the proposed approach has three stages: COVID-19 infected region segmentation, infection map generation, and COVID-19 detection. The CXR image is the input to the trained E-D CNN and the network's probabilistic prediction is used to generate infection maps. The generated infection maps are used for COVID-19 detection.}
    \label{fig:approach}
\end{figure*}

The rest of the paper is organized as follows. In Section \ref{dataset}, we introduce the benchmark QaTa-COV19 dataset. Our novel human-machine collaborative approach for the ground-truth annotation is explained in Section \ref{HMCollab}. Next, the details of COVID-19 infected region segmentation, and the infection map generation and COVID-19 detection are presented  in Sections \ref{pneumonia-segmentation} and \ref{heatmap-detection}, respectively\footnote{The live demo of the proposed approach is implemented on \href{http://qatacov.live/}{http://qatacov.live/}}. The experimental setup and results with the benchmark dataset are reported in Section \ref{experimental-setup} and \ref{experimental-results}, respectively. Finally, we conclude the paper in Section \ref{conclusion}.

\section{Materials and Methodology}
The proposed approach in this study is composed of three main phases: 1) training the state-of-the-art deep models for COVID-19 infected region segmentation using the ground-truth segmentation masks, 2) infection map generation from the trained segmentation networks, and 3) COVID-19 detection as it can be depicted in Fig. \ref{fig:approach}. In this section, we first detail the creation of the benchmark QaTa-COV19 dataset. Then, the proposed approach for collaborative human-machine ground-truth generation is introduced. 

\subsection{The Benchmark QaTa-COV19 Dataset}\label{dataset}

 \begin{figure}[b!]
    \centering
    \includegraphics[width=0.40\textwidth]{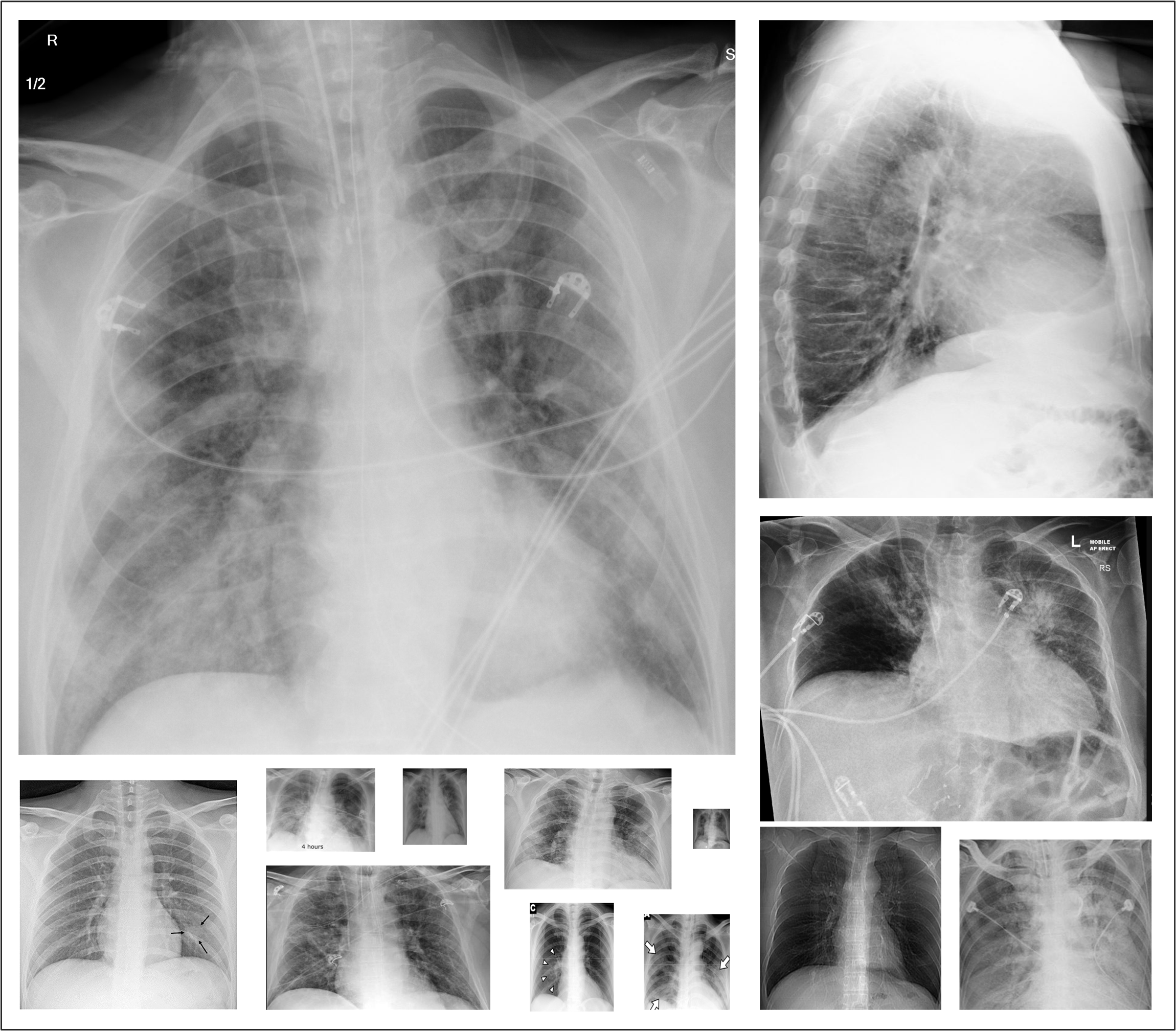}
    \caption{The COVID-19 CXR samples from the benchmark QaTa-COV19 dataset.}
    \label{fig:dataset}
\end{figure}

The researchers of Qatar University and Tampere University have compiled the largest COVID-19 dataset up to date with nearly $120$K CXR images: \textit{QaTa-COV19} including \textbf{$2951$} COVID-19 CXRs. To create QaTa-COV19, we have utilized several publicly available, scattered, and different format datasets and repositories. Therefore, the collected images from the datasets had some duplicate, over-exposed and low-quality images that were identified and removed in the pre-processing stage. Consequently, the COVID-19 CXRs are from different publicly available sources resulting in high intra-class dissimilarity as depicted in Fig. \ref{fig:dataset}. The image sources of COVID-19 and control group CXRs are detailed as follows:

\textbf{COVID-19 CXRs:} BIMCV-COVID19+ \cite{vaya2020bimcv} is the largest publicly available dataset with $2473$ COVID-19 positive CXR images. The CXR images of BIMCV-COVID19+ dataset were recorded with computed radiography (CR) and digital X-ray (DX) machines. Hannover Medical School and Institute for Diagnostic and Interventional Radiology \cite{covidimage} released $183$ CXR images of COVID-19 patients. A total of $959$ CXR images are from public repositories: Italian Society of Medical and Interventional Radiology (SIRM), GitHub, and Kaggle \cite{coviddatabase, cohen2020covid, radiodatabase, chestimaging, haghanifar2020covid}. As mentioned earlier, any duplication and low-quality images are removed since COVID-19 CXR images are collected from different public datasets and repositories. In this study, a total of $2951$ COVID-19 CXRs are gathered from the aforementioned datasets. Therefore, COVID-19 CXRs are of different age, group, gender, and ethnicity.

\begin{figure*}[t!]
    \centering
    \includegraphics[width=0.95\textwidth]{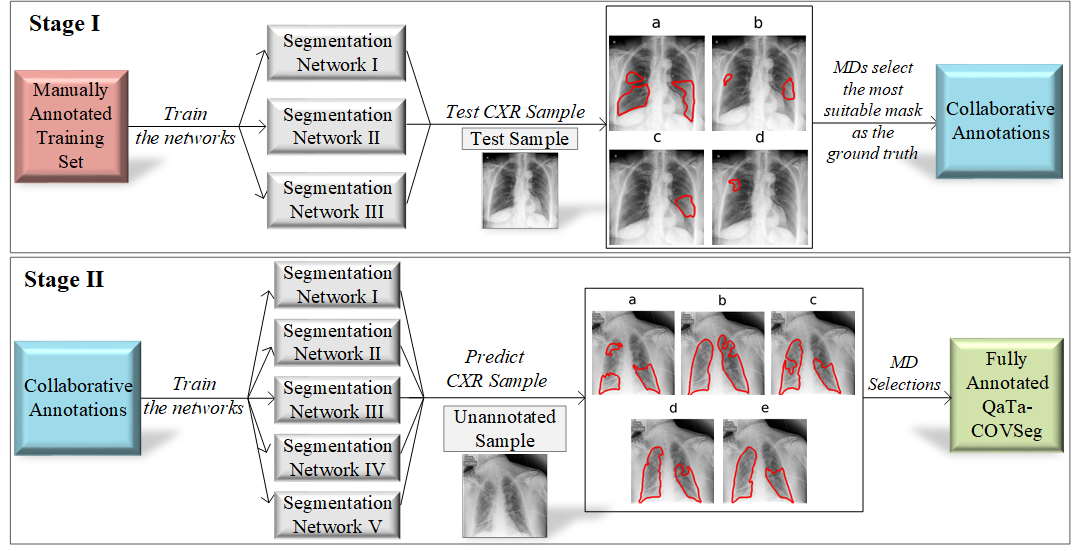}
    \caption{The two stages of the human-machine collaborative approach. Stage I: A subset of CXR images with manually drawn segmentation masks are used to train three different deep networks in a 5-fold cross-validation scheme. The manually drawn ground-truth (a), and the three predictions (b, c, d) are blindly shown to MDs, and they select the best ground-truth mask. Stage II: Five deep networks are trained over the best segmentation masks selected. Then, they are used to produce the segmentation masks for the rest of the CXR dataset (a, b, c, d, e), which are shown to MDs.}
    \label{fig:stages_GT}
\end{figure*}

\textbf{Control Group CXRs:} In this study, we have considered two control groups in the experimental evaluation. Group-I consists of only normal (healthy) CXRs with a smaller number of images compared to the second group. RSNA pneumonia detection challenge dataset \cite{RSNA} is comprised of about $29.7$K CXR images, where $8851$ images are normal. All CXRs in the dataset are in DICOM format, a popularly used format for medical imaging. Padchest dataset \cite{bustos2020padchest} consists of $160,868$ CXR images from $67,625$ patients, where $37,871$ images are from normal class. The images are evaluated and reported by radiologists at Hospital Sun Juan in Spain during $2009-2017$. The dataset includes six different position views of CXR and additional information regarding image acquisition and patient demography. Paul Mooney \cite{kermany2018identifying} has released an X-ray dataset of $5863$ CXR images from a total of $5856$ patients, where $1583$ images are from normal class. The data is collected from pediatric patients aging one to five years old at Guangzhou Women and Children’s Medical Center, Guangzhou. The dataset in \cite{indiana} consists of $7470$ CXR images and the corresponding radiologist reports from the Indiana Network for Patient Care, where a total of $1343$ frontal CXR samples are labeled as normal. In \cite{chinaUS}, there are $80$ normal CXRs from the tuberculosis control program of the Department of Health and Human Services of Montgomery County and $326$ normal CXRs from Shenzhen Hospital. In this study, a total of $12,544$ normal CXRs are included in control Group-I from the aforementioned datasets. On the other hand, Group-II consists of \textbf{$116,365$} CXRs from $15$ different classes. ChestX-ray14 \cite{wang2017chestx} consists of $112,120$ CXRs with normal and $14$ different thoracic disease images, which are atelectasis, cardiomegaly, effusion, inﬁltration, mass, nodule, pneumonia, pneumothorax, consolidation, edema, emphysema, fibrosis, PT, hernia, and normal (no findings). Additionally, from the pediatric patients \cite{kermany2018identifying},  $2760$ bacterial and $1485$ viral pneumonia CXRs are included in Group-II.

\subsection{Collaborative Human-Machine Ground-Truth Annotation}\label{HMCollab}
Recent developments in the machine and deep learning techniques led to state-of-the-art performance in many computer vision (CV) tasks, such as image classification, object detection, and image segmentation. However, supervised DL methods require a huge amount of annotated data. Otherwise, the limited amount of data degrades the performance of the deep network structures since their generalization capability depends on the availability of large datasets. Nevertheless, to produce ground-truth segmentation masks, pixel-accurate image segmentation by human experts can be a cumbersome and highly subjective task even for moderate size datasets. 

In order to overcome this challenge, in this study, we propose a novel collaborative human-machine approach to accurately produce the ground-truth segmentation masks for infected regions directly from the CXR images. The proposed approach is performed in two main stages. First, a group of expert MDs manually segment the infected regions of a subset of ($500$ in our case) CXR images. Then, several segmentation networks that are inspired by the U-Net \cite{ronneberger2015u} structure with a 5-fold cross-validation scheme, are trained over the initial ground-truth masks. For each fold, the segmentation masks of the test samples are predicted by the networks. The network predicted masks along with the initial (MD drawn) ground-truth masks, and original CXR image are assessed by the MDs, and the best segmentation mask among them is selected. Steps of Stage-I are illustrated in Fig. \ref{fig:stages_GT} (top). At the end of the first stage, collaboratively annotated ground-truth masks for the subset of CXR images are formed, and they are obviously superior to the initial manually drawn masks since they are selected by the MDs. An interesting observation in this stage was that MDs preferred the machine-generated masks over the manually drawn masks in the first stage in three out of five cases.

In the second stage five deep networks, inspired by U-Net \cite{ronneberger2015u}, UNet++ \cite{zhou2018unet++}, and DLA \cite{yu2018deep} architectures are trained over the collaborative masks, which were formed in Stage-I. The trained segmentation networks are used to predict the segmentation masks of the rest of the data, which is around $2400$ unannotated COVID-19 images. Among the five predictions, the expert MDs select the best one as the ground-truth or deny all if none was found successful. For the latter case, MDs were asked to draw the ground-truth masks manually. However, we notice that this was indeed a minority case that included less than $5$\% of unannotated data. The steps of Stage-II are shown in Fig. \ref{fig:stages_GT} (bottom). As a result, the ground-truth masks for $2951$ COVID-19 CXR images are gathered to construct the benchmark QaTa-COV19 dataset. The proposed approach does not only save valuable human labor time, but it also improves the quality and reliability of the masks by reducing the subjectivity with Stage-II verification step.

\subsection{COVID-19 Infected Region Segmentation}\label{pneumonia-segmentation}
Segmentation of COVID-19 infection is the first step of our proposed approach as depicted in Fig. \ref{fig:approach}. Once the ground-truth annotation for QaTa-COV19 benchmark dataset is formed as explained in the previous section, we perform infected region segmentation extensively with $24$ different network configurations. We have used three different segmentation models: U-Net, UNet++ and DLA, with four different encoder structures: CheXNet, DenseNet-121, Inception-v3 and ResNet-50, and frozen \& not frozen encoder weight configurations. 

\subsubsection{Segmentation Models}
We have tried distinct segmentation model structures starting from shallow to deep structures with varied configurations as follows:

\begin{itemize}
    \item \textbf{U-Net} \cite{ronneberger2015u} is an outperforming network for medical image segmentation applications with a u-shaped architecture as the encoder part is symmetric with respect to its decoder part. Therefore, this unique decoder structure with many feature channels allows the network to carry the information through its latest layers. 
    \item \textbf{UNet++} \cite{zhou2018unet++} has further developed the decoder structure of U-Net by connecting the encoder to the decoder with the nested dense convolutional blocks. This way, the bridge between the encoder and decoder parts are more firmly knit; thus, the information can be transferred to its final layers more intensively compared to the classic U-Net. 
    \item \textbf{DLA} \cite{yu2018deep} investigates the connecting bridges between the encoder and decoder, and proposes a way to fuse the semantic and spatial information with dense layers, which are progressively aggregated by iterative merging to deeper and larger scales.  
\end{itemize}

\subsubsection{Encoder Selections for Segmentation Models}
In this study, we use several deep CNNs to form the encoder part of the above-mentioned segmentation models as follows:

\begin{itemize}
    \item \textbf{DenseNet-121} \cite{huang2017densely} is a deep network with $121$ layers, each with additional input nodes connecting all the layers directly with each other. Therefore, the maximum information flow through the network is satisfied.
    \item \textbf{CheXNet} \cite{rajpurkar2017chexnet} is based on the architecture of DenseNet-121, which is trained over the ChestX-ray14 dataset \cite{wang2017chestx} to detect pneumonia cases from CXR images. In \cite{rajpurkar2017chexnet}, DenseNet-121 is initialized with the ImageNet weights and fine-tuned over $100$K CXR images resulting from the state-of-the-art results on the ChestX-ray14 dataset with a better performance compared to the conclusions of radiologists. 
    \item \textbf{Inception-v3} \cite{szegedy2016rethinking} achieves state-of-the-art results with much less computational complexity compared to its deep competitors by factorizing the convolutions and pruning the dimensions inside the network. Despite the less complexity, it preserves a higher performance.
    \item \textbf{ResNet-50} \cite{he2016deep} introduces a deep residual learning framework that forces the desired mapping of the input to a residual mapping. It is possible to achieve this goal by the shortcut connections on the stacked layers. These connections enable to merge the input and output of the stacked layers by addition operations; therefore, the problem of gradient vanishing is prevented.
\end{itemize}
We perform transfer learning on the encoder side of the segmentation models by initializing the layers with the ImageNet weights, except for CheXNet which is pre-trained on the ChestX-ray14 dataset. We tried two configurations, in the first we freeze the encoder layers while in the second, they are allowed to vary. 

\subsubsection{Hybrid Loss Function}
In this study, we have performed training the segmentation networks with a hybrid loss function by combining \textit{focal loss} \cite{lin2017focal} with \textit{dice loss} \cite{milletari2016v} to achieve a better segmentation performance. We use focal loss since COVID-19 infected region segmentation is an imbalanced problem: the number of background pixels is superior to the foreground's. Let the ground-truth segmentation mask be $\mathbf{Y}$, where each pixel class label is defined as $y$, and the network prediction as $\hat{y}$. We define the pixel class probabilities as for the positive class $P(y=1)=p$, and for the negative class $P(y=0)=1-p$. On the other hand, the network prediction probabilities are modeled by the logistic function using the sigmoid curve as,
\begin{equation}
    P(\hat{y}=1)=\frac{1}{1+e^{-z}}=q
\end{equation}
\begin{equation}
    P(\hat{y}=0)=1-\frac{1}{1+e^{-z}}=1-q
\end{equation}
where $z$ is some function of the input CXR image $\mathbf{X}$. Then, we define the cross-entropy (CE) loss as follows:
\begin{equation}
    CE(p, q) = - p\log q - (1-p)\log(1-q).
\end{equation}
A common solution to address the class imbalance problem is to add a weighting factor $\alpha \in [0,1]$ for the positive class, and $1-\alpha$ for the negative class, which defines the balanced cross-entropy (BCE) loss as,
\begin{equation}
    BCE(p, q) = -\alpha p \log q - (1 - \alpha) (1-p)\log(1-q).
\end{equation}
In this way, the importance of positive and negative samples are balanced. However, adding the $\alpha$ factor does not solve the issue for the large class imbalance scenario. This is because the network cannot distinguish outliers (hard samples) and inliers (easy samples) with the BCE loss. To overcome this drawback, focal loss \cite{lin2017focal} proposes to set focusing parameter $\gamma\geq0$ in order to down-weight the loss of easy samples that occur with small errors; so that the model can be forced to learn hard negative samples. The focal (F) loss is defined as,
\begin{equation}
    F(p, q) = -\alpha (1-q)^\gamma p \log q - (1 - \alpha) q^\gamma (1-p)\log(1-q).
\end{equation}
where F loss is equivalent to BCE loss when $\gamma = 0$. In our experimental setup, we use the default setting as $\alpha=0.25$, and $\gamma=2$ for all the networks. To achieve a good segmentation performance, we combined focal loss with dice loss, which is based on the dice coefficient (DC) defined as follows:
\begin{equation}
    DC = \frac{2 |\mathbf{Y} \cap \mathbf{\hat{Y}}|}{|\mathbf{Y}| \cup |\mathbf{\hat{Y}}|}
\end{equation}
where $\mathbf{\hat{Y}}$ is the predicted segmentation mask of the network. Hence, the DC can be interpreted as a dice (D) loss as follows:
\begin{equation}
    D(p, q) = 1 - \frac{2 \sum p_{h, w} q_{h, w}}{\sum p_{h,w} + \sum q_{h, w}}
\end{equation}
where $h$ and $w$ are the height and width of the ground-truth and prediction masks $\mathbf{Y}$ and $\mathbf{\hat{Y}}$, respectively. Finally, we combined D and F losses by summation to achieve the so-called hybrid loss function for the segmentation networks.

\subsection{Infection Map Generation and COVID-19 Detection}\label{heatmap-detection}
\begin{figure}[t!]
    \centering
    \includegraphics[width=0.48\textwidth]{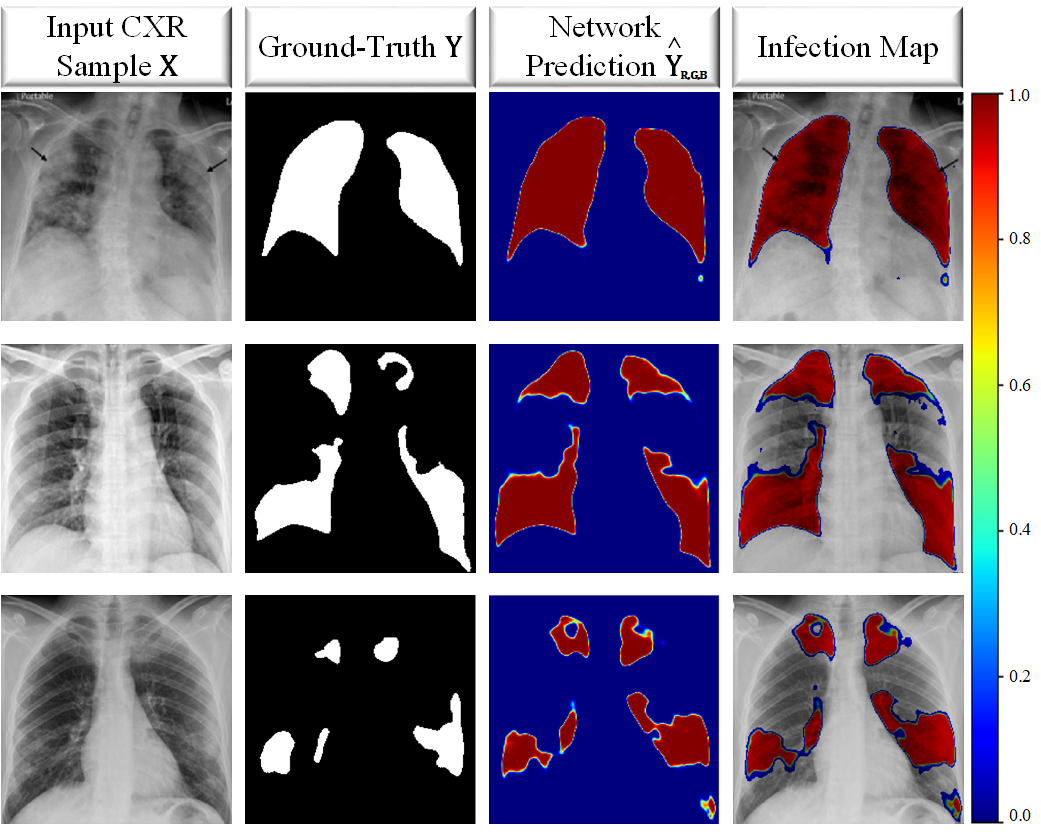}
    \caption{The three COVID-19 CXR test samples, $\mathbf{X}$ with the corresponding ground-truth masks, $\mathbf{Y}$. The color-coded network predictions, $\mathbf{\hat{Y}_{R, G, B}}$ are reflected translucent onto the $\mathbf{X}$ to generate an infection map on the lungs, where $\mathbf{\hat{Y}}>0$.}
    \label{fig:heat_map}
\end{figure}

Having the training set of COVID-19 CXR images via the collaborative human-machine approach explained in Section \ref{dataset}, we train the aforementioned segmentation networks to produce infection maps. After training the segmentation networks, we feed each test CXR sample $\mathbf{X}$ into the trained network. Then, we obtain the network prediction mask $\mathbf{\hat{Y}}$, which is used to generate an infection map that is a measure of infected region probabilities on the input $\mathbf{X}$. Each pixel in $\mathbf{\hat{Y}}$ is defined as $\mathbf{\hat{Y}_{h,w}} \in [0,1]$, where $h$ and $w$ represent the size of the image. We then apply an RGB-based color transform, i.e., the jet color scale to obtain the RGB version of the prediction mask, $\mathbf{\hat{Y}_{\text{R,G,B}}}$ as shown in Fig. \ref{fig:heat_map} for a pseudo-colored probability measure visualization. The infection map is generated as a reflection of the network prediction $\mathbf{\hat{Y}_{\text{R,G,B}}}$ onto the CXR image $\mathbf{X}$. Hence, for visualization, we form the imposed image by concatenating the hue and saturation components of $\mathbf{\hat{Y}_{\text{H,S,V}}}$, and value component of $\mathbf{X_{\text{H,S,V}}}$. Finally, the imposed image is converted back to RGB domain. In the infection map, we do not show the pixels/regions with zero probabilities for a better visualization effect. This way, the infected regions, where $\mathbf{\hat{Y}}>0$ are shown translucent as in Fig \ref{fig:heat_map}. 

Along with the infection map generation, which already provides localization and segmentation of COVID-19 infection, COVID-19 detection can easily be performed using the proposed approach. The detection of COVID-19 is performed based on the predictions of the trained segmentation networks. Accordingly, a test sample is classified as COVID-19 class if $\mathbf{\hat{Y}\geq0.5}$ at any pixel location.  

\section{Experimental Results}
In this section, first, the experimental setup is presented. Then, both numerical and visual results are reported with an extensive set of comparative evaluations over the benchmark QaTa-COV19 dataset. Finally, visual comparative evaluations are presented between the infection maps and the activation maps extracted from state-of-the-art deep models.

\subsection{Experimental Setup}\label{experimental-setup}
Quantitative evaluations for the proposed approach are performed for both COVID-19 infected region segmentation and COVID-19 detection. COVID-19 infected region segmentation is evaluated on a pixel-level, where we consider the foreground (infected region) as the positive class, and background as the negative class. For COVID-19 detection, the performance is computed per CXR sample, and we consider COVID-19 as the positive class and the control group as the negative class. Overall, elements of the confusion matrix are formed as follows: true positive (TP): the number of correctly detected positive class members, true negative (TN): the number of correctly detected negative class samples, false positive (FP): the number of misclassified negative class members, and false negative (FN): the number of misclassified positive class samples. The standard performance evaluation metrics are defined as follows:
\begin{equation}
    Sensitivity = \frac{TP}{TP+FN}
\end{equation}
where \textit{sensitivity} (or Recall) is the rate of correctly detected positive samples in the positive class samples,
\begin{equation}
    Specificity = \frac{TN}{TN+FP}
\end{equation} 
where \textit{specificity} is the ratio of accurately detected negative class samples to all negative class samples,
\begin{equation}
    Precision = \frac{TP}{TP+FP}
\end{equation}
where \textit{precision} is the rate of correctly classified positive class samples among all the members classified as positive samples,
\begin{equation}
    Accuracy = \frac{TP+TN}{TP+TN+FP+FN}
\end{equation}

\begin{table*}[t!]
\centering
\caption{Average performance metrics (\%) for \textit{COVID-19 infected region segmentation} computed on the Group-I test (unseen) set from $5$-folds with three state-of-the-art segmentation models, four encoder architectures, and weight initializations. The initialized encoder layers are set to \textit{frozen} (\cmark) and \textit{not frozen} (\xmark) states during the investigation.}
\resizebox{\textwidth}{!}{
\begin{tabular}{|c|c|c|ccccccc|} \hline
\begin{tabular}[c]{@{}c@{}}Model\end{tabular}  & Encoder & \begin{tabular}[c]{@{}c@{}}Encoder \\ Layers\end{tabular} & Sensitivity & Specificity & Precision & F1-Score & F2-Score & Accuracy & AUC \\ \hline \hline

\multirow{8}{*}{\large{U-Net}} & CheXNet & \cmark & $81.20 \pm 1.6\times 10^{-4}$ & $99.55 \pm 5\times 10^{-6}$ & $83.78 \pm 2.6\times 10^{-5}$ & $82.47 \pm 2.7\times 10^{-5}$ & $81.70 \pm  2.7\times 10^{-5}$ & $99.03 \pm 6.9\times 10^{-6}$ &  $99.19 \pm 6.3\times 10^{-6}$\\

 & CheXNet & \xmark & \cellcolor[gray]{0.95}$82.23 \pm 1.6\times 10^{-4}$
 & \cellcolor[gray]{0.95}$99.56 \pm 5\times 10^{-6}$ & \cellcolor[gray]{0.95}$84.54 \pm 2.5\times 10^{-5}$ & \cellcolor[gray]{0.95}$83.34 \pm 2.6\times 10^{-5}$ & \cellcolor[gray]{0.95}$82.66 \pm 2.7\times 10^{-5}$ & \cellcolor[gray]{0.95}$99.08 \pm 6.7\times 10^{-6}$ & \cellcolor[gray]{0.95}$99.18 \pm 6.3\times 10^{-6}$ \\
 
 & DenseNet-121 & \cmark & $82.29 \pm 1.6\times 10^{-4}$ & $99.61 \pm 4\times 10^{-6}$ & $86.02 \pm 2.4\times 10^{-5}$ & $84.11 \pm 2.6\times 10^{-5}$ & $83.01 \pm 2.6\times 10^{-5}$ & $99.13 \pm 6.5\times 10^{-6}$ & $\textbf{99.35} \pm 5.6\times 10^{-6}$\\
 
 & DenseNet-121 & \xmark &  \cellcolor[gray]{0.95}$\textbf{84.00} \pm 1.5\times 10^{-4}$ & \cellcolor[gray]{0.95}$99.66 \pm 4\times 10^{-6}$ & \cellcolor[gray]{0.95}$87.77 \pm 2.3\times 10^{-5}$ & \cellcolor[gray]{0.95}$\textbf{85.81} \pm 2.5\times 10^{-5}$ & \cellcolor[gray]{0.95}$\textbf{84.71} \pm 2.5\times 10^{-5}$ & \cellcolor[gray]{0.95}$\textbf{99.22} \pm 6.2\times 10^{-6}$ & \cellcolor[gray]{0.95}$99.19 \pm 6.3\times 10^{-6}$ \\
 
 & Inception-v3 & \cmark  & $80.42 \pm 1.7\times 10^{-4}$ & $99.59 \pm 5\times 10^{-6}$ & $84.94 \pm 2.5\times 10^{-5}$ & $82.62 \pm 2.7\times 10^{-5}$ & $81.28 \pm 2.7\times 10^{-5}$ & $99.05 \pm 6.8\times 10^{-6}$ & $99.20 \pm 6.3\times 10^{-6}$ \\
 
 & Inception-v3 & \xmark  &\cellcolor[gray]{0.95}$82.34 \pm 1.6\times 10^{-4}$ &\cellcolor[gray]{0.95}$\textbf{99.70} \pm 4\times 10^{-6}$ & \cellcolor[gray]{0.95}$\textbf{88.87} \pm 2.2\times 10^{-5}$ & \cellcolor[gray]{0.95}$85.43 \pm 2.5\times 10^{-5}$ & \cellcolor[gray]{0.95}$83.54 \pm 2.6\times 10^{-5}$ & \cellcolor[gray]{0.95}$99.21 \pm 6.2\times 10^{-6}$ & \cellcolor[gray]{0.95}$98.82 \pm 7.6\times 10^{-6}$ \\
 
 & ResNet-50 & \cmark & $81.43 \pm 1.6\times 10^{-4}$ & $99.62 \pm 4\times 10^{-6}$ & $86.07 \pm 2.4\times 10^{-5}$ & $83.67 \pm 2.6\times 10^{-5}$ & $82.31 \pm 2.6\times 10^{-5}$ & $99.11 \pm 6.6\times 10^{-6}$
 & $99.30 \pm 5.9\times 10^{-6}$\\
 
 & ResNet-50 & \xmark  & \cellcolor[gray]{0.95}$79.90 \pm 1.7\times 10^{-4}$ & \cellcolor[gray]{0.95}$\textbf{99.70} \pm 4\times 10^{-6}$ & \cellcolor[gray]{0.95}$88.64 \pm 2.2\times 10^{-5}$ & \cellcolor[gray]{0.95}$83.89 \pm 2.6\times 10^{-5}$ & \cellcolor[gray]{0.95}$81.43 \pm 2.7\times 10^{-5}$ & \cellcolor[gray]{0.95}$99.15 \pm 6.5\times 10^{-6}$ & \cellcolor[gray]{0.95}$98.98 \pm 7.1\times 10^{-6}$ \\
 
 \hline \hline
 
 \multirow{8}{*}{\large{UNet++}} & CheXNet & \cmark  & $80.29 \pm 1.7\times 10^{-4}$ & $99.59 \pm 5\times 10^{-6}$ & $85.19 \pm 2.5\times 10^{-5}$ & $82.64 \pm 2.7\times 10^{-5}$ & $81.21 \pm 2.7\times 10^{-5}$ & $99.05 \pm 6.8\times 10^{-6}$ & $99.01 \pm 7\times 10^{-6}$ \\

 & CheXNet & \xmark  & \cellcolor[gray]{0.95}$81.45 \pm 1.6\times 10^{-4}$ & \cellcolor[gray]{0.95}$99.60 \pm 5\times 10^{-6}$ & \cellcolor[gray]{0.95}$85.60 \pm 2.5\times 10^{-5}$& \cellcolor[gray]{0.95}$83.47 \pm 2.6\times 10^{-5}$ &\cellcolor[gray]{0.95}$82.24 \pm 2.7\times 10^{-5}$ & \cellcolor[gray]{0.95}$99.09 \pm 6.7\times 10^{-6}$ & \cellcolor[gray]{0.95}$99.01 \pm 7\times 10^{-6}$ \\
 
 & DenseNet-121 & \cmark  & $82.38 \pm 1.6\times 10^{-4}$ & $99.61 \pm 4\times 10^{-6}$ & $85.99 \pm 2.4\times 10^{-5}$ & $84.14 \pm 2.6\times 10^{-5}$ & $83.08 \pm 2.6\times 10^{-5}$ & $99.13 \pm 6.5\times 10^{-6}$ & $99.19 \pm 6.3\times 10^{-6}$ \\
 
 & DenseNet-121 & \xmark  & \cellcolor[gray]{0.95}$82.36 \pm 1.6\times 10^{-4}$ & \cellcolor[gray]{0.95}$\textbf{99.68} \pm 4\times 10^{-6}$ & \cellcolor[gray]{0.95}$\textbf{88.07} \pm 2.3\times 10^{-5}$ & \cellcolor[gray]{0.95}$85.08 \pm 2.5\times 10^{-5}$ & \cellcolor[gray]{0.95}$83.42 \pm 2.6\times 10^{-5}$ & \cellcolor[gray]{0.95}$99.19 \pm 6.3\times 10^{-6}$ & \cellcolor[gray]{0.95}$\textbf{99.30} \pm 5.9\times 10^{-6}$ \\
 
 & Inception-v3 & \cmark  & $82.87 \pm 1.6\times 10^{-4}$ & $99.57 \pm 5\times 10^{-6}$ & $84.83 \pm 2.5\times 10^{-5}$ & $83.81 \pm 2.6\times 10^{-5}$ & $83.24 \pm 2.6\times 10^{-5}$ & $99.10 \pm 6.6\times 10^{-6}$ & $99.21 \pm 6.2\times 10^{-6}$ \\
 
 & Inception-v3 & \xmark  & \cellcolor[gray]{0.95}$\textbf{83.49} \pm 1.6\times 10^{-4}$ & \cellcolor[gray]{0.95}$99.66 \pm 4\times 10^{-6}$ & \cellcolor[gray]{0.95}$87.60 \pm 2.3\times 10^{-5}$ & \cellcolor[gray]{0.95}$\textbf{85.45} \pm 2.5\times 10^{-5}$ & \cellcolor[gray]{0.95}$\textbf{84.22} \pm 2.6\times 10^{-5}$ & \cellcolor[gray]{0.95}$\textbf{99.20} \pm 6.3\times 10^{-6}$ & \cellcolor[gray]{0.95}$99.18 \pm 6.3\times 10^{-6}$ \\
 
 & ResNet-50 & \cmark  & $82.07 \pm 1.6\times 10^{-4}$ & $99.59 \pm 5\times 10^{-6}$ & $85.41 \pm 2.5\times 10^{-5}$ & $83.71 \pm 2.6\times 10^{-5}$ & $82.72 \pm 2.7\times 10^{-5}$ & $99.10 \pm 6.6\times 10^{-6}$ & $99.15 \pm 6.5\times 10^{-6}$ \\
 
 & ResNet-50 & \xmark  & \cellcolor[gray]{0.95}$82.64 \pm 1.6\times 10^{-4}$ & \cellcolor[gray]{0.95}$99.62 \pm 4\times 10^{-6}$ & \cellcolor[gray]{0.95}$86.52 \pm 2.4\times 10^{-5}$ & \cellcolor[gray]{0.95}$84.45 \pm 2.5\times 10^{-5}$ & \cellcolor[gray]{0.95}$83.33 \pm 2.6\times 10^{-5}$ & \cellcolor[gray]{0.95}$99.14 \pm 6.5\times 10^{-6}$ & \cellcolor[gray]{0.95}$99.27 \pm 6\times 10^{-6}$ \\
 
 \hline \hline
 
 \multirow{8}{*}{\large{DLA}} & CheXNet & \cmark  & $79.99 \pm 1.7\times 10^{-4}$ & $99.61 \pm 4\times 10^{-6}$ & $85.57 \pm 2.5\times 10^{-5}$ & $82.66 \pm 2.7\times 10^{-5}$ & $81.04 \pm 2.8\times 10^{-5}$ & $99.06 \pm 6.8\times 10^{-6}$ & $99.12 \pm 6.6\times 10^{-6}$ \\

 & CheXNet & \xmark  & \cellcolor[gray]{0.95}$82.84 \pm 1.6\times 10^{-4}$ & \cellcolor[gray]{0.95}$99.56 \pm 5\times 10^{-6}$ & \cellcolor[gray]{0.95}$84.63 \pm 2.5\times 10^{-5}$ & \cellcolor[gray]{0.95}$83.71 \pm 2.6\times 10^{-5}$ & \cellcolor[gray]{0.95}$83.19 \pm 2.6\times 10^{-5}$ & \cellcolor[gray]{0.95}$99.09 \pm 6.7\times 10^{-6}$ & \cellcolor[gray]{0.95}$99.17 \pm 6.4\times 10^{-6}$ \\
 
 & DenseNet-121 & \cmark  & $82.48 \pm 1.6\times 10^{-4}$ & $99.62 \pm 4\times 10^{-6}$ & $86.40 \pm 2.4\times 10^{-5}$ & $84.36 \pm 2.6\times 10^{-5}$ & $83.21 \pm 2.6\times 10^{-5}$ & $99.14 \pm 6.5\times 10^{-6}$ & $99.16 \pm 6.4\times 10^{-6}$ \\
 
 & DenseNet-121 & \xmark  & \cellcolor[gray]{0.95}$82.84 \pm 1.6\times 10^{-4}$ & \cellcolor[gray]{0.95}$99.56 \pm 5\times 10^{-6}$ & \cellcolor[gray]{0.95}$84.63 \pm 2.5\times 10^{-5}$ & \cellcolor[gray]{0.95}$83.71 \pm 2.6\times 10^{-5}$ & \cellcolor[gray]{0.95}$83.19 \pm 2.6\times 10^{-5}$ & \cellcolor[gray]{0.95}$99.09 \pm 6.7\times 10^{-6}$ & \cellcolor[gray]{0.95}$99.17 \pm 6.4\times 10^{-6}$ \\
 
 & Inception-v3 & \cmark  & $80.28 \pm 1.7\times 10^{-4}$ & $99.63 \pm 4\times 10^{-6}$ & $86.43 \pm 2.4\times 10^{-5}$ & $83.19 \pm 2.6\times 10^{-5}$ & $81.41 \pm 2.7\times 10^{-5}$ & $99.09 \pm 6.7\times 10^{-6}$ & $99.02 \pm 6.9\times 10^{-6}$ \\
 
 & Inception-v3 & \xmark  & \cellcolor[gray]{0.95}$\textbf{83.44} \pm 1.6\times 10^{-4}$ & \cellcolor[gray]{0.95}$\textbf{99.68} \pm 4\times 10^{-6}$ & \cellcolor[gray]{0.95}$\textbf{88.18} \pm  2.3\times 10^{-5}$& \cellcolor[gray]{0.95}$\textbf{85.73} \pm 2.5\times 10^{-5}$ & \cellcolor[gray]{0.95}$\textbf{84.34} \pm 2.6\times 10^{-5}$ & \cellcolor[gray]{0.95}$\textbf{99.22} \pm 6.2\times 10^{-6}$ & \cellcolor[gray]{0.95}$\textbf{99.29} \pm 5.9\times 10^{-6}$ \\
 
 & ResNet-50 & \cmark  & $81.26 \pm 1.6\times 10^{-4}$ & $99.63 \pm 4\times 10^{-6}$ & $86.48 \pm 2.4\times 10^{-5}$ & $83.78 \pm 2.6\times 10^{-5}$ & $82.25 \pm 2.7\times 10^{-5}$ & $99.12 \pm 6.6\times 10^{-6}$ & $99.08 \pm 6.7\times 10^{-6}$ \\
 
 & ResNet-50 & \xmark  & \cellcolor[gray]{0.95}$82.07 \pm 1.6\times 10^{-4}$ & \cellcolor[gray]{0.95}$99.65 \pm 4\times 10^{-6}$ & \cellcolor[gray]{0.95}$86.99 \pm 2.4\times 10^{-5}$ & \cellcolor[gray]{0.95}$84.45 \pm 2.5\times 10^{-5}$ & \cellcolor[gray]{0.95}$83.00 \pm 2.6\times 10^{-5}$ & \cellcolor[gray]{0.95}$99.15 \pm 6.5\times 10^{-6}$ & \cellcolor[gray]{0.95}$99.31 \pm 5.8\times 10^{-6}$ \\
\hline
\end{tabular}}
\label{table-segmentation-results}
\end{table*}

\noindent where \textit{accuracy} is the ratio of correctly classified elements among all the data,
\begin{equation}
    F({\beta}) = (1 + \beta ^ 2) \frac{(Precision \times Sensitivity)}{\beta ^ 2 \times Precision + Sensitivity}
\end{equation}
where $F$-score is defined by the weighting parameter $\beta$. The \textit{F1}-Score is calculated with $\beta = 1$, which is the harmonic average of \textit{precision} and \textit{sensitivity}. The \textit{F2}-score is calculated with $\beta=2$, which emphasizes FN minimization over FPs. The main objective of both COVID-19 segmentation and detection is to maximize \textit{sensitivity} with a reasonable \textit{specificity} in order to minimize FP COVID-19 cases or pixels. Equivalently, maximized \textit{F2}-score is targeted with an acceptable \textit{F1}-Score value. The performances with their $95$\% confidence interval (CI) for both COVID-19 infected region segmentation and detection are given in Tables \ref{table-segmentation-results} and \ref{table-detection-results}, respectively. The range of values can be calculated for each performance as follows: 
\begin{equation}
r = \pm z\sqrt{metric(1-metric)/N},
\end{equation}
where $z$ is the level of significance, $metric$ is any performance evaluation metric, and $N$ is the number of samples. Accordingly, $z$ is set to $1.96$ for $95$\% CI. 

We have implemented the deep networks with Tensorflow library \cite{abadi2016tensorflow} using Python on NVidia ® GeForce RTX 2080 Ti GPU card. For training, Adam optimizer \cite{kingma2014adam} is used with the default momentum parameters, $\beta1 = 0.9$ and $\beta2 = 0.999$ using the aforementioned hybrid loss function. The segmentation networks are trained with $50$-epochs with a learning rate of $\alpha = 10^{-4}$ and a batch size of $32$. 

For comparing the computed infection maps, the activation maps are computed as follows: the encoder structures of the segmentation networks are trained for the classification task with a modification at the output layer by adding $2$-neurons for the number of total classes. The activation maps extracted from the classification models are then compared with the infection maps of the segmentation models. The classification networks, CheXNet, DenseNet-121, Inception-v3 and ResNet-50 are fine-tuned using categorical cross-entropy as loss function with $10$ epochs and a learning rate of $\alpha = 10^{-5}$, which is a sufficient setting to prevent over-fitting, based on our previous study \cite{ahishali2020comparative}. Other settings of the classifiers are kept the same with the segmentation models. 

\subsection{Experimental Results}\label{experimental-results}
The experiments are carried out for both COVID-19 infected region segmentation and COVID-19 detection. We extensively tested the benchmark QaTa-COV19 dataset using three different state-of-the-art segmentation networks with four different encoder options for the initial dataset consisting of control Group-I. We also investigated the effect of frozen encoder weights on the performance. On the other hand, the leading model is selected and evaluated on the extended dataset, which includes more negative samples with the control Group-II.

\begin{table}[b!]
\centering
\caption{Number of CXR samples in control Group-I per fold before and after data augmentation.}
\begin{tabular}{|c|c|c|c|c|}
\hline
\rowcolor[gray]{.90}Data & \begin{tabular}[c]{@{}c@{}}Number of \\ Samples\end{tabular} & \begin{tabular}[c]{@{}c@{}}Training \\ Samples\end{tabular} & \begin{tabular}[c]{@{}c@{}}Augmented \\ Training Samples\end{tabular} & \begin{tabular}[c]{@{}c@{}}Test \\ Samples\end{tabular} \\ \hline \hline

\begin{tabular}[c]{@{}c@{}}COVID-19\end{tabular} & $2951$ & $2361$ & $10035$ & $590$ \\  \hline

\rowcolor[gray]{.95}\begin{tabular}[c]{@{}c@{}}Group-I\end{tabular} & $12544$ & $10035$ & $10035$ & $2509$ \\ \hline \hline

Total & $15495$ & $12396$ & $\textbf{20070}$ & $\textbf{3099}$ \\ \hline
\end{tabular}
\label{numberofsamples}
\end{table}

\subsubsection{Group-I Experiments} We have evaluated the networks in a stratified 5-fold cross-validation scheme with a ratio of $80\%$ training to $20\%$ test (unseen folds) over the benchmark QaTa-COV19 dataset. The input CXR images are resized to $224\times 224$ pixels. Table \ref{numberofsamples} shows the number of CXRs per fold in the dataset. Since the two classes are imbalanced, we have applied data augmentation in order to balance the classes. Therefore, COVID-19 samples are augmented up to the same number of samples as the control Group-I in the training set for each fold. The data augmentation is performed using Image Data Generator in Keras: the CXR samples are augmented by randomly shifting them both vertically and horizontally by $10\%$ and randomly rotating them in a range of $10$ degrees. 

\begin{table*}[t!]
\centering
\caption{Average \textit{COVID-19 detection} performance results (\%) computed from $5$-folds over the Group-I test (unseen) set with three network models, four encoder architectures, and weight initializations. The initialized encoder layers are set to \textit{frozen} (\cmark) and \textit{not frozen} (\xmark) states during the investigation.}
\resizebox{\textwidth}{!}{
\begin{tabular}{c|c|c|cccccc|}
\cline{2-9}
 & \multicolumn{1}{c|}{Encoder} & \begin{tabular}[c]{@{}c@{}}Encoder \\ Layers\end{tabular} & \multicolumn{1}{c}{Sensitivity} & \multicolumn{1}{c}{Specificity} & \multicolumn{1}{c}{Precision} & \multicolumn{1}{c}{F1-Score} & \multicolumn{1}{c}{F2-Score} & Accuracy \\ \hline
 
\multicolumn{1}{|c|}{\multirow{9}{*}{{\rotatebox[origin=c]{90}{\large{U-Net}}}}} & CheXNet & \cmark & $97.56 \pm 0.0056$ & $91.10 \pm 0.0050$ & $72.07 \pm 0.0071$ & $82.90 \pm 0.0059$ & $91.11 \pm 0.0045$ & $92.33 \pm 0.0042$ \\
\multicolumn{1}{|c|}{} & CheXNet & \xmark & \cellcolor[gray]{0.95}$97.97 \pm 0.0051$  & \cellcolor[gray]{0.95}$92.74 \pm 0.0045$ & \cellcolor[gray]{0.95}$76.04 \pm 0.0067$  & \cellcolor[gray]{0.95}$85.62 \pm 0.0055$ & \cellcolor[gray]{0.95}$92.62 \pm 0.0041$ & \cellcolor[gray]{0.95}$93.73 \pm 0.0038$ \\
\multicolumn{1}{|c|}{} & DenseNet-121 & \cmark & $98.07 \pm 0.0050$ & $94.66 \pm 0.0039$ & $81.20 \pm 0.0062$ & $88.84 \pm 0.0050$ & $94.16 \pm 0.0037$ & $95.31 \pm 0.0033$ \\
\multicolumn{1}{|c|}{} & DenseNet-121 & \xmark & \cellcolor[gray]{0.95}$\textbf{98.37} \pm 0.0046$ & \cellcolor[gray]{0.95}$98.05 \pm 0.0024$ & \cellcolor[gray]{0.95}$92.25 \pm 0.0042$ & \cellcolor[gray]{0.95}$95.21 \pm 0.0034$ & \cellcolor[gray]{0.95}$\textbf{97.08} \pm 0.0027$ & \cellcolor[gray]{0.95}$98.12 \pm 0.0021$ \\
\multicolumn{1}{|c|}{} & Inception-v3 & \cmark & $97.93 \pm 0.0051$ & $90.00 \pm 0.0052$ & $69.74 \pm 0.0072$ & $81.47 \pm 0.0061$ & $90.61 \pm 0.0046$ & $91.51 \pm 0.0044$ \\
\multicolumn{1}{|c|}{} & Inception-v3 & \xmark & \cellcolor[gray]{0.95}$97.22 \pm 0.0059$ & \cellcolor[gray]{0.95}$\textbf{98.37} \pm 0.0022$ & \cellcolor[gray]{0.95}$\textbf{93.33} \pm 0.0039$ & \cellcolor[gray]{0.95}$\textbf{95.24} \pm 0.0034$  & \cellcolor[gray]{0.95}$96.42 \pm 0.0029$ & \cellcolor[gray]{0.95}$\textbf{98.15} \pm 0.0021$ \\
\multicolumn{1}{|c|}{} & ResNet-50 & \cmark & $98.24 \pm 0.0047$
 & $93.88 \pm 0.0042$ & $79.06 \pm 0.0064$ & $87.61 \pm 0.0052$ & $93.69 \pm 0.0038$ & $94.71 \pm 0.0035$ \\
\multicolumn{1}{|c|}{} & ResNet-50 & \xmark & \cellcolor[gray]{0.95}$96.37 \pm 0.0067$  & \cellcolor[gray]{0.95}$97.82 \pm 0.0026$ & \cellcolor[gray]{0.95}$91.21 \pm 0.0045$ & \cellcolor[gray]{0.95}$93.72 \pm 0.0038$ & \cellcolor[gray]{0.95}$95.30 \pm 0.0033$ & \cellcolor[gray]{0.95}$97.54 \pm 0.0024$ \\ \hline

\multicolumn{1}{|c|}{\multirow{9}{*}{{\rotatebox[origin=c]{90}{\large{UNet++}}}}} & CheXNet & \cmark  & $97.80 \pm 0.0053$ & $91.70 \pm 0.0048$ & $73.49 \pm 0.0069$ & $83.92 \pm 0.0058$ & $91.73 \pm 0.0043$ & $92.86 \pm 0.0041$ \\
\multicolumn{1}{|c|}{} & CheXNet & \xmark & \cellcolor[gray]{0.95}$97.49 \pm 0.0056$  & \cellcolor[gray]{0.95}$93.65 \pm 0.0043$ & \cellcolor[gray]{0.95}$78.33 \pm 0.0065$ &  \cellcolor[gray]{0.95}$86.87 \pm 0.0053$  & \cellcolor[gray]{0.95}$92.94 \pm 0.0040$ & \cellcolor[gray]{0.95}$94.39 \pm 0.0036$ \\
\multicolumn{1}{|c|}{} & DenseNet-121 & \cmark  & $97.70 \pm 0.0054$ & $94.81 \pm 0.0039$ & $81.58 \pm 0.0061$ & $88.91 \pm 0.0049$ & $93.98 \pm 0.0037$ & $95.36 \pm 0.0033$ \\
\multicolumn{1}{|c|}{} & DenseNet-121 & \xmark & \cellcolor[gray]{0.95}$96.51 \pm 0.0066$ & \cellcolor[gray]{0.95}$\textbf{99.16} \pm 0.0016$ & \cellcolor[gray]{0.95}$\textbf{96.44} \pm 0.0029$ & \cellcolor[gray]{0.95}$\textbf{96.48} \pm 0.0029$ & \cellcolor[gray]{0.95}$\textbf{96.50} \pm 0.0029$ & \cellcolor[gray]{0.95}$\textbf{98.66} \pm 0.0018$ \\
\multicolumn{1}{|c|}{} & Inception-v3 & \cmark  & $\textbf{98.31} \pm 0.0047$ & $90.54 \pm 0.0051$ & $70.96 \pm 0.0071$ & $82.43 \pm 0.0060$ & $91.27 \pm 0.0044$ & $92.02 \pm 0.0043$ \\
\multicolumn{1}{|c|}{} & Inception-v3 & \xmark & \cellcolor[gray]{0.95}$96.92 \pm 0.0061$  & \cellcolor[gray]{0.95}$98.37 \pm 0.0022$ & \cellcolor[gray]{0.95}$93.34 \pm 0.0039$ & \cellcolor[gray]{0.95}$95.10 \pm 0.0034$ & \cellcolor[gray]{0.95}$96.18 \pm 0.0030$ & \cellcolor[gray]{0.95}$98.10 \pm 0.0021$ \\
\multicolumn{1}{|c|}{} & ResNet-50 & \cmark  & $97.80 \pm 0.0053$ & $93.39 \pm 0.0043$ & $77.69 \pm 0.0066$ & $86.59 \pm 0.0054$ & $92.98 \pm 0.0040$ & $94.23 \pm 0.0037$ \\
\multicolumn{1}{|c|}{} & ResNet-50 & \xmark & \cellcolor[gray]{0.95}$96.78 \pm 0.0064$ & \cellcolor[gray]{0.95}$97.43 \pm 0.0028$ & \cellcolor[gray]{0.95}$89.87 \pm 0.0048$ & \cellcolor[gray]{0.95}$93.20 \pm 0.0040$  & \cellcolor[gray]{0.95}$95.31 \pm 0.0033$  & \cellcolor[gray]{0.95}$97.31 \pm 0.0025$ \\ \hline

\multicolumn{1}{|c|}{\multirow{9}{*}{{\rotatebox[origin=c]{90}{\large{DLA}}}}} & CheXNet & \cmark  & $97.46 \pm 0.0057$ & $92.47 \pm 0.0046$ & $75.27 \pm 0.0068$ & $84.94 \pm 0.0056$ & $92.03 \pm 0.0043$ & $93.42 \pm 0.0039$ \\
\multicolumn{1}{|c|}{} & CheXNet & \xmark & \cellcolor[gray]{0.95}$97.32 \pm 0.0058$ & \cellcolor[gray]{0.95}$94.93 \pm 0.0038$ & \cellcolor[gray]{0.95}$81.87 \pm 0.0061$ & \cellcolor[gray]{0.95}$88.93 \pm 0.0049$ & \cellcolor[gray]{0.95}$93.78 \pm 0.0038$ & \cellcolor[gray]{0.95}$95.39 \pm 0.0033$ \\
\multicolumn{1}{|c|}{} & DenseNet-121 & \cmark  & $97.36 \pm 0.0058$ & $95.66 \pm 0.0036$ & $84.08 \pm 0.0058$ & $90.23 \pm 0.0047$ & $94.38 \pm 0.0036$ & $95.99 \pm 0.0031$ \\
\multicolumn{1}{|c|}{} & DenseNet-121 & \xmark & \cellcolor[gray]{0.95}$97.09 \pm 0.0061$ & \cellcolor[gray]{0.95}$99.07 \pm 0.0017$ & \cellcolor[gray]{0.95}$96.08 \pm 0.0031$ & \cellcolor[gray]{0.95}$\textbf{96.58} \pm 0.0029$ &\cellcolor[gray]{0.95}$\textbf{96.88} \pm 0.0027$ & \cellcolor[gray]{0.95}$\textbf{98.69} \pm 0.0018$ \\
\multicolumn{1}{|c|}{} & Inception-v3 & \cmark  & $96.92 \pm 0.0062$ & $93.24 \pm 0.0044$ & $77.13 \pm 0.0066$  & $85.90 \pm 0.0055$ & $92.19 \pm 0.0042$ & $93.94 \pm 0.0040$ \\
\multicolumn{1}{|c|}{} & Inception-v3 & \xmark & \cellcolor[gray]{0.95}$96.71 \pm 0.0064$ & \cellcolor[gray]{0.95}$\textbf{99.13} \pm 0.0016$ & \cellcolor[gray]{0.95}$\textbf{96.32} \pm 0.0030$ & \cellcolor[gray]{0.95}$96.52 \pm 0.0029$ & \cellcolor[gray]{0.95}$96.63 \pm 0.0028$ & \cellcolor[gray]{0.95}$98.67 \pm 0.0018$ \\
\multicolumn{1}{|c|}{} & ResNet-50 & \cmark  & $\textbf{97.49} \pm 0.0056$ & $95.30 \pm 0.0037$ & $82.98 \pm 0.0059$ & $89.65 \pm 0.0048$ & $94.20 \pm 0.0037$ & $95.71 \pm 0.0032$ \\
\multicolumn{1}{|c|}{} & ResNet-50 & \xmark & \cellcolor[gray]{0.95}$96.17 \pm 0.0069$ & \cellcolor[gray]{0.95}$98.15 \pm 0.0024$ & \cellcolor[gray]{0.95}$92.44 \pm 0.0042$ & \cellcolor[gray]{0.95}$94.27 \pm 0.0037$ & \cellcolor[gray]{0.95}$95.40 \pm 0.0033$ & \cellcolor[gray]{0.95}$97.77 \pm 0.0023$ \\ \hline
\end{tabular}}
\label{table-detection-results}
\end{table*} 

\begin{figure}[b!]
    \centering
    \includegraphics[width=0.4\textwidth]{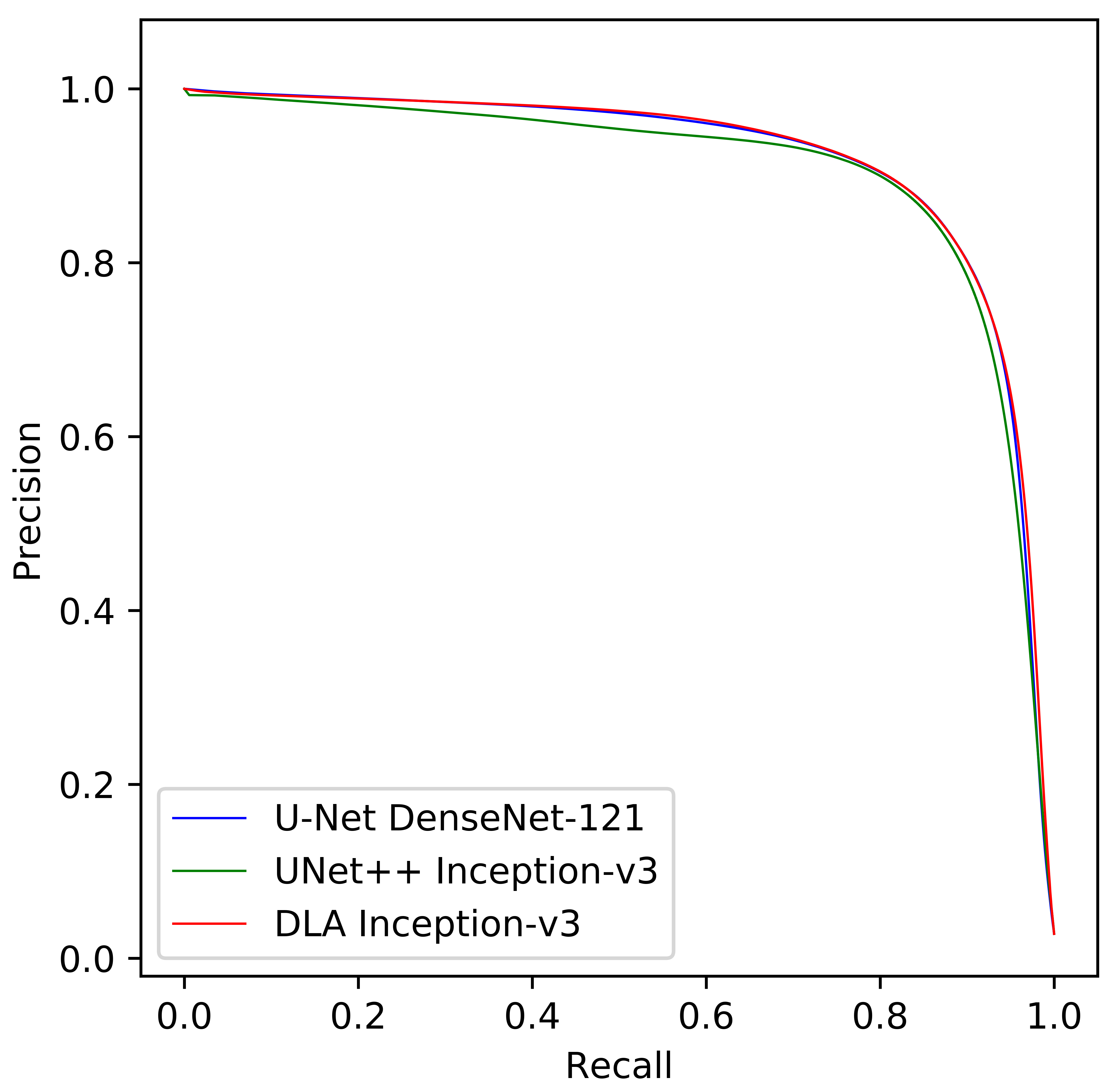}
    \caption{The Precision-Recall curves of the three leading models all with the not frozen encoder layers setting.}
    \label{fig:precision-recall}
\end{figure} 

\noindent After shifting and rotating the images, blank sections are filled using the \textit{nearest} mode.

The performance of the segmentation models for COVID-19 infected region segmentation are presented in Table \ref{table-segmentation-results}. Each model structure is evaluated with two configurations: \textit{frozen} and \textit{not frozen} encoder layers. We have used transfer learning on the encoder layers with ImageNet weights, except for the CheXNet model, which is pre-trained on the ChestX-ray14 dataset. The evaluation of the models with frozen encoder layers is also important since this process can lead to a better convergence and improved performance. However, as the results show, better performance is obtained when the network continues to learn on the encoder layers as well. For each model, we have observed that two encoders: DenseNet-121 and Inception-v3 are the top-performing ones for the infected region segmentation task. The U-Net model with DenseNet-121 encoder holds the leading performance by $84$\% sensitivity, $85.81$\% F1-Score, and $84.71$\% F2-Score. DenseNet-121 produces better results compared to other encoder types since it can preserve the information coming from earlier layers through the output by concatenating the feature maps from each dense layer. However, in the other segmentation models, Inception-v3 outperforms the other encoder types. The presented segmentation performances are obtained by setting the threshold value to $0.5$ to compute the segmentation mask from the network probabilities. The Precision-Recall curves are plotted in Fig. \ref{fig:precision-recall} by varying this threshold value.

The performances of the segmentation models for COVID-19 detection are presented in Table \ref{table-detection-results}. All the models are evaluated by stratified a 5-fold cross-validation scheme, and the table shows the averaged results of these folds. The most crucial metric here is the sensitivity since missing any patient with COVID-19 is critical. In fact, the results indicate the robustness of the model as the proposed approach can achieve high sensitivity levels of $98.37$\% with a $97.08$\% F2-Score. Additionally, the proposed approach achieves an elegant specificity of $99.16$\%, indicating a significantly low false alarm rate. It can be observed from Table \ref{table-detection-results} that DenseNet-121 encoder with the not frozen encoder layer setting gives the most promising results among the others. The confusion matrices, accumulated on each fold's test set, are presented in Table \ref{CMs}. The highest sensitivity in COVID-19 detection is achieved by the U-Net DenseNet-121 model (Table \ref{CMa}). Accordingly, the U-Net DenseNet-121 model only misses $48$ COVID-19 patients out of $2951$. On the other hand, the highest specificity is achieved by UNet++ DenseNet-121 model (Table \ref{CMb}). The UNet++ model only misses a minor part of the control class with $105$ samples out of $12544$. 

\begin{table}[t!]
\centering
\caption{Cumulative confusioFn matrices of COVID-19 detection by the best performing U-Net and UNet++ models with DenseNet-121 encoder.}
\begin{subtable}{.48\textwidth}
\centering
\caption{U-Net DenseNet-121}
\small
    \begin{tabular}{|c|c|c|c|}
\hline
\multicolumn{2}{|c|}{\multirow{2}{*}{\textbf{U-Net}}} & \multicolumn{2}{c|}{Predicted} \\ \cline{3-4} 
\multicolumn{2}{|c|}{} & \multicolumn{1}{c|}{Group-I} & \multicolumn{1}{c|}{COVID-19} \\ \hline
\multirow{2}{*}{\begin{tabular}[c]{@{}c@{}}Ground\\ Truth\end{tabular}} & Group-I & $12300$ & $244$ \\ \cline{2-4} 
 & COVID-19 & $48$ & $2903$ \\ \hline
\end{tabular}
\label{CMa}
\end{subtable}

\bigskip
\noindent
\begin{subtable}{.48\textwidth}
\centering
\caption{UNet++ DenseNet-121}
\small
    \begin{tabular}{|c|c|c|c|}
\hline
\multicolumn{2}{|c|}{\multirow{2}{*}{\textbf{UNet++}}} & \multicolumn{2}{c|}{Predicted} \\ \cline{3-4} 
\multicolumn{2}{|c|}{} & \multicolumn{1}{c|}{Group-I} & \multicolumn{1}{c|}{COVID-19} \\ \hline
\multirow{2}{*}{\begin{tabular}[c]{@{}c@{}}Ground\\ Truth\end{tabular}} & Group-I & $12439$ & $105$ \\ \cline{2-4} 
 & COVID-19 & $103$ & $2848$ \\ \hline
\end{tabular}
\label{CMb}
\end{subtable}
\label{CMs}
\end{table}

\subsubsection{Group-II Experiments} We have selected the leading model from the Group-I experiments as U-Net with not frozen DenseNet-121 encoder setting. In Group-II experiments, we have gathered around $120$K CXRs. The CXRs from the ChestX-ray14 dataset \cite{wang2017chestx} are already divided into train and test sets. Accordingly, we have randomly separated the train and test sets of COVID-19, viral pneumonia, and bacterial pneumonia CXRs by keeping the same train/test ratio as in ChestX-ray14 \cite{wang2017chestx}. Table \ref{numberofsamplesGROUPII} shows the number of training and test samples of the Group-II experiments. Additonally, we have applied augmentation to data except for ChestX-ray14 samples with the same set-up as in the Group-I experiments. In these experiments, we do not perform any cross-validation since ChestX-ray14 has predefined training and test sets.

\begin{table}[ht!]
\centering
\caption{Number of CXR samples in control Group-II before and after data augmentation.}
\begin{tabular}{|c|c|c|c|c|}
\hline
\rowcolor[gray]{.90}Data & \begin{tabular}[c]{@{}c@{}}Training \\ Samples\end{tabular} & \begin{tabular}[c]{@{}c@{}}Augmented\end{tabular} & \begin{tabular}[c]{@{}c@{}}Augmented \\ Training Samples\end{tabular} & \begin{tabular}[c]{@{}c@{}}Test \\ Samples\end{tabular} \\ \hline \hline

\begin{tabular}[c]{@{}c@{}}COVID-19 \\ \end{tabular} & $2078$ & \cmark & $10,000$ & $873$ \\  \hline

\rowcolor[gray]{.95}\begin{tabular}[c]{@{}c@{}}Bacterial \\ Pneumonia\end{tabular} & $2130$ & \cmark & $5000$ &  $630$\\  \hline

\begin{tabular}[c]{@{}c@{}}ChestX-ray14\end{tabular} & $86,524$ & \xmark & $86,524$ & $25,596$ \\ \hline

\rowcolor[gray]{.95}\begin{tabular}[c]{@{}c@{}}Viral \\ Pneumonia\end{tabular} & $1146$ & \cmark & $5000$ & $339$ \\ \hline \hline

Total & 91,878 &  & $\textbf{106,524}$ & $\textbf{27,438}$ \\ \hline
\end{tabular}
\label{numberofsamplesGROUPII}
\end{table}

The  performance  of  the  U-Net  model  for  COVID-19 infected region segmentation and detection is presented in Table \ref{table-Group-IIresults}. The model achieved a segmentation performance by $81.72$\% sensitivity and $83.20$\% F1-Score. In comparison to initial experiments with the control Group-I data, the model can still achieve an elegant segmentation performance even with numerous samples in the test set. On the other hand, the COVID-19 detection performance with $27,438$ CXR images is very successful by $94.96$\% sensitivity, $99.88$\% specificity, and $96.40$\% precision. This indicates a very low false alarm rate of only $0.12$\%. Table \ref{CMGROUP-II} shows the confusion matrix on the test set. Accordingly, the model only misses $44$ COVID-19 samples. In the control Group-II, only $31$ CXR samples are missed, which is a minor section in $26,565$ negative samples. The results show that the leading model is still robust on the extended data, where it consists of $15$ different classes with $14$ thoracic diseases and normal samples.

\begin{table}[t!]
\centering
\caption{COVID-19 \textit{infected region segmentation} and \textit{detection} results (\%) computed on the Group-II test set from the U-Net model with DenseNet-121 encoder.}
\small
\begin{tabular}{c|c|c|c|}
\cline{2-4} & \multicolumn{1}{c|}{\begin{tabular}[c|]{@{}c@{}}Performance \\ Metrics\end{tabular}} & \multicolumn{1}{c|}{\begin{tabular}[c|]{@{}c@{}}Infected Region \\ Segmentation\end{tabular}} & \multicolumn{1}{c|}{Detection}  \\ \hline
\multicolumn{1}{|c|}{\multirow{6}{*}{{\rotatebox[origin=c]{90}{\begin{tabular}[c]{@{}c@{}}\large{U-Net} \\ DenseNet-121\end{tabular} }}}} & \cellcolor[gray]{0.95}Sensitivity  & \cellcolor[gray]{0.95}$81.72$  &  \cellcolor[gray]{0.95}$94.96$ \\
\multicolumn{1}{|c|}{} & Specificity & $99.93$ & $99.88$\\
\multicolumn{1}{|c|}{} & \cellcolor[gray]{0.95}Precision & \cellcolor[gray]{0.95}$84.74$ & \cellcolor[gray]{0.95}$96.40$ \\
\multicolumn{1}{|c|}{} & F1-Score & $83.20$ & $95.67$ \\ 
\multicolumn{1}{|c|}{} & \cellcolor[gray]{0.95}F2-Score  & \cellcolor[gray]{0.95}$82.31$ & \cellcolor[gray]{0.95}$95.24$ \\ 
\multicolumn{1}{|c|}{} & Accuracy & $99.85$ & $99.73$ \\ \hline
\end{tabular}
\label{table-Group-IIresults}
\end{table}

\begin{table}[t!]
\centering
\caption{Cumulative confusion matrices of COVID-19 detection by the best performing U-Net model with DenseNet-121 encoder.}
\centering
\small
    \begin{tabular}{|c|c|c|c|}
\hline
\multicolumn{2}{|c|}{\multirow{2}{*}{\textbf{U-Net}}} & \multicolumn{2}{c|}{Predicted} \\ \cline{3-4} 
\multicolumn{2}{|c|}{} & \multicolumn{1}{c|}{Group-II} & \multicolumn{1}{c|}{COVID-19} \\ \hline
\multirow{2}{*}{\begin{tabular}[c]{@{}c@{}}Ground\\ Truth\end{tabular}} & Group-II & $26,534$ & $31$ \\ \cline{2-4} 
 & COVID-19 & $44$ & $829$ \\ \hline
\end{tabular}
\label{CMGROUP-II}
\end{table}

\begin{figure*}[t!]
    \centering
    \includegraphics[width=1\textwidth]{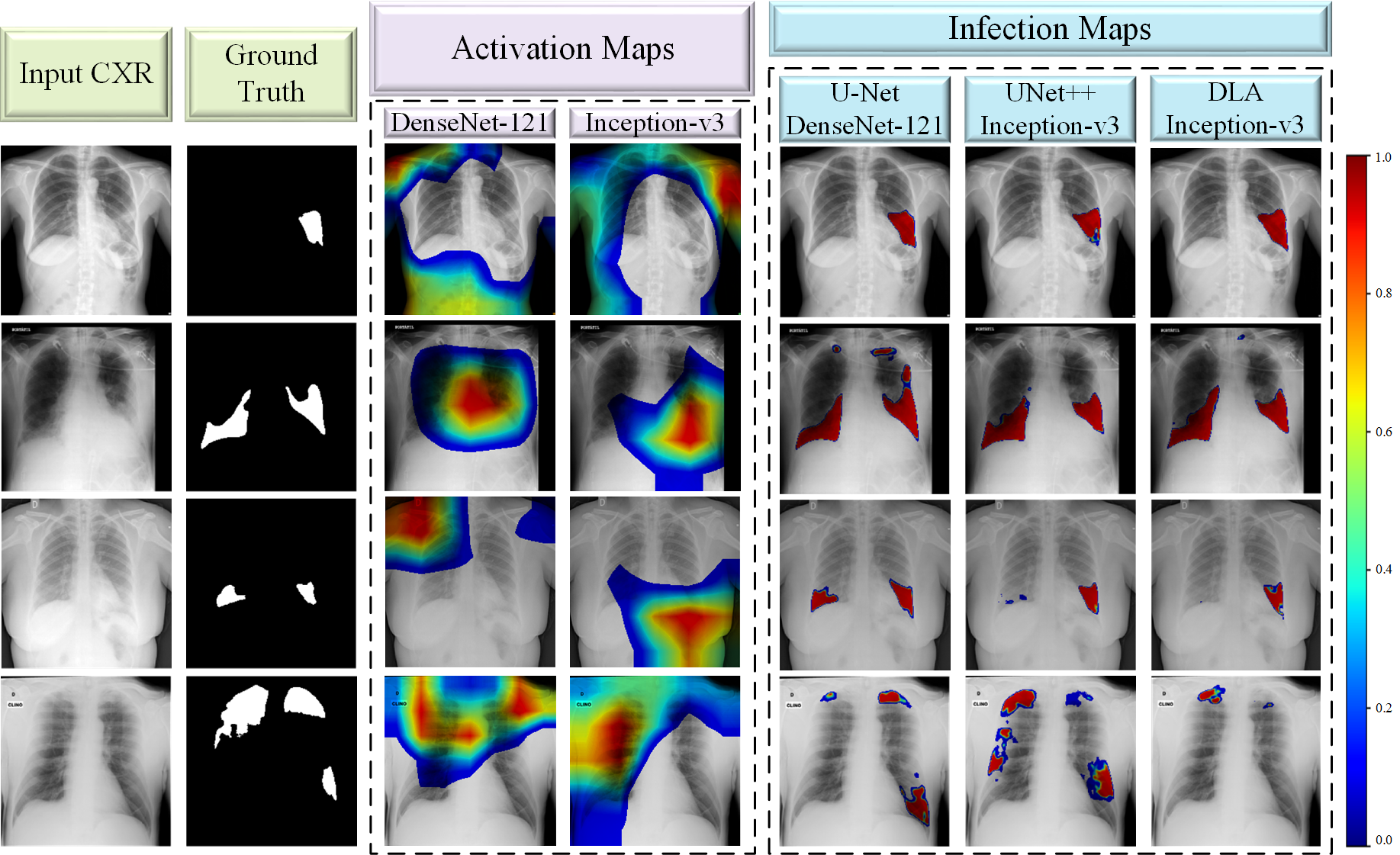}
    \caption{Several CXR images with their corresponding ground-truth masks. The activation maps extracted from the classification models are presented in the middle block. The last block is the generated infection maps from the segmentation models. It is evident that the infection maps yield a superior localization of COVID-19 infection compared to activation maps.}
    \label{fig:comparision}
\end{figure*}

\subsection{Infection vs Activation Maps}\label{heat-activation}
Several studies \cite{yeh2020cascaded, oh2020deep, ozturk2020automated} propose to localize COVID-19 from CXRs by extracting activation maps from the deep classification models trained for COVID-19 detection. Despite the simplicity of the idea, there are many limitations of this approach. First of all, without any infected region segmentation ground-truth masks, the network can only produce a rough localization, and the extracted activation maps may entirely fail to localize COVID-19 infection.

In this study, we check the reliability of our proposed COVID-19 detection approach by comparing it with DL models trained for the classification task. In order to achieve this objective, we compare the infection map and activation map of CXR images, which are generated from the segmentation and classification networks, respectively. Therefore, we have trained the encoder structures of the segmentation networks, which are CheXNet, DenseNet-121, Inception-v3, and ResNet-50 to perform COVID-19 classification task. We have extracted activation maps from these trained models by the Gradient-weighted Class Activation Mapping (Grad-CAM) approach proposed in \cite{selvaraju2017grad}. The localization Grad-CAM $L^c_{\text{Grad-CAM}} \in \mathbb{R}^{h \times w}$ of height $h$ and width $w$ for class $c$ is calculated by the gradient of $m^c$ before the softmax with respect to the convolutional layer's feature maps $A^k$ as $\frac{\partial{m^c}}{\partial{A^k}}$. The gradients are passed through from the global average pooling during back-propagation;
\begin{equation}
    \alpha^c_k = \frac{1}{Z}\sum_i\sum_j\frac{\partial{m^c}}{\partial{A^k}},
\end{equation}
where $\alpha$ is the weight that shows the important feature map $k$ from $A$ for a target class $c$. Then, the linear combination is performed following by ReLU to obtain the Grad-CAM;
\begin{equation}
    L^c_{\text{Grad-CAM}}=ReLU(\sum_k \alpha_k^c A^k).
\end{equation}

Despite their elegant performance, activation maps extracted from deep classification networks are not suitable for localizing COVID-19 infection as depicted in Fig \ref{fig:comparision}. In fact, infections found by the activation maps are highly irrelevant indicating false locations outside of the lung areas. On the other hand, infection maps can generate a highly accurate location with an elegant severity grading of COVID-19 infection. The proposed infection maps can conveniently be used by medical experts for an enhanced assessment of the disease. Real-time implementation of the infection maps will obviously speed up the detection process, can also monitor the progression of COVID-19 infection in the lungs.

\begin{table}[t!]
\centering
\caption{The number of trainable and non-trainable parameters of the models with their inference time (ms) per sample. The initialized encoder layers are set to frozen (\cmark) \textit{or} not frozen (\xmark).}
\begin{tabular}{c|c|c|ccc|}
\cline{2-6}
 & \multicolumn{1}{c|}{Encoder} & \begin{tabular}[c]{@{}c@{}}Encoder \\ Layers \end{tabular} & \multicolumn{1}{c}{Trainable} & \multicolumn{1}{c}{Non-Trainable}  & \begin{tabular}[c|]{@{}c@{}}Time \\ (ms)\end{tabular} \\ \hline
 
\multicolumn{1}{|c|}{\multirow{8}{*}{{\rotatebox[origin=c]{90}{\large{U-Net}}}}} & CheXNet & \cmark & $5.19$M & $6.96$M & $2.56$ \\
\multicolumn{1}{|c|}{} & DenseNet-121 & \cmark & \cellcolor[gray]{0.95}$5.19$M  & \cellcolor[gray]{0.95}$6.96$M & \cellcolor[gray]{0.95}$2.58$ \\
\multicolumn{1}{|c|}{} & Inception-v3 & \cmark & $8.15$M & $21.79$M & $\textbf{2.53}$\\
\multicolumn{1}{|c|}{} & ResNet-50 & \cmark & \cellcolor[gray]{0.95}$9.06$M & \cellcolor[gray]{0.95}$23.50$M & \cellcolor[gray]{0.95}$2.54$ \\ 
\multicolumn{1}{|c|}{} & CheXNet & \cmark & $12.06$M & $85.63$K & $2.62$ \\ 
\multicolumn{1}{|c|}{} & DenseNet-121 & \xmark & \cellcolor[gray]{0.95}$12.06$M & \cellcolor[gray]{0.95}$85.63$K & \cellcolor[gray]{0.95}$2.58$ \\
\multicolumn{1}{|c|}{} & Inception-v3 & \xmark & $29.9$M & $36.42$K & $2.61$ \\
\multicolumn{1}{|c|}{} & ResNet-50 & \xmark & \cellcolor[gray]{0.95}$32.51$M & \cellcolor[gray]{0.95}$47.56$K & \cellcolor[gray]{0.95}$2.64$ \\ \hline \hline

\multicolumn{1}{|c|}{\multirow{8}{*}{{\rotatebox[origin=c]{90}{\large{UNet++}}}}} & CheXNet & \cmark  & $7.53$M & $6.96$M & $5.17$ \\
\multicolumn{1}{|c|}{} & DenseNet-121 & \cmark  & \cellcolor[gray]{0.95}$7.53$M & \cellcolor[gray]{0.95}$6.96$M & \cellcolor[gray]{0.95}$5.10$ \\
\multicolumn{1}{|c|}{} & Inception-v3 & \cmark  & $8.68$M & $21.79$M & $5.32$ \\
\multicolumn{1}{|c|}{} & ResNet-50 & \cmark & \cellcolor[gray]{0.95}$10.88$M & \cellcolor[gray]{0.95}$23.51$M & \cellcolor[gray]{0.95}$5.58$ \\
\multicolumn{1}{|c|}{} & CheXNet & \xmark & $14.40$M & $88.45$K & $5.24$ \\
\multicolumn{1}{|c|}{} & DenseNet-121 & \xmark & \cellcolor[gray]{0.95}$14.40$M & \cellcolor[gray]{0.95}$88.45$K & \cellcolor[gray]{0.95}$5.25$ \\
\multicolumn{1}{|c|}{} & Inception-v3 & \xmark & $30.43$M & $39.23$K & $5.32$ \\
\multicolumn{1}{|c|}{} & ResNet-50 & \xmark & \cellcolor[gray]{0.95}$34.34$M  & \cellcolor[gray]{0.95}$50.37$K & \cellcolor[gray]{0.95}$5.46$ \\ \hline \hline

\multicolumn{1}{|c|}{\multirow{8}{*}{{\rotatebox[origin=c]{90}{\large{DLA}}}}} & CheXNet & \cmark  & $6.27$M & $6.96$M  & $4.65$ \\
\multicolumn{1}{|c|}{} & DenseNet-121 & \cmark  & \cellcolor[gray]{0.95}$6.27$M & \cellcolor[gray]{0.95}$6.96$M & \cellcolor[gray]{0.95}$4.63$ \\
\multicolumn{1}{|c|}{} & Inception-v3 & \cmark  & $7.20$M & $21.79$M & $4.70$\\
\multicolumn{1}{|c|}{} & ResNet-50 & \cmark & \cellcolor[gray]{0.95}$8.74$M & \cellcolor[gray]{0.95}$23.51$M & \cellcolor[gray]{0.95}$4.90$ \\
\multicolumn{1}{|c|}{} & CheXNet & \xmark & $13.15$M & $88.45$K & $4.63$ \\
\multicolumn{1}{|c|}{} & DenseNet-121 & \xmark & \cellcolor[gray]{0.95}$13.15$M & \cellcolor[gray]{0.95}$88.45$K & \cellcolor[gray]{0.95}$4.65$ \\
\multicolumn{1}{|c|}{} & Inception-v3 & \xmark & $28.96$M & $39.23$K & $4.72$ \\
\multicolumn{1}{|c|}{} & ResNet-50 & \xmark & \cellcolor[gray]{0.95}$32.2$M
 & \cellcolor[gray]{0.95}$50.37$K & \cellcolor[gray]{0.95}$4.90$ \\ \hline
\end{tabular}
\label{table-parameters}
\end{table}  

\subsection{Computational Complexity Analysis}
In this section, we present the computational times of the networks and their number of trainable \& non-trainable parameters. Table \ref{table-parameters} shows the elapsed time in milliseconds (ms) during the inference step for each network used in the experiments. The results in the table represent the running time per sample. It can be observed from the table that the U-Net model is the fastest among the others due to its shallow structure. The fastest network is U-Net Inception-v3 with frozen encoder layers taking up $2.53$ ms. On the other hand, the slowest model is UNet++ structure since it has the largest number of trainable parameters. The most computationally demanding model is UNet++ ResNet-50 with frozen encoder layers, which takes $5.58$ ms. We, therefore, conclude that all models can be used as real-time clinical applications. 

\section{Conclusions}\label{conclusion}
The immediate and accurate detection of highly infectious COVID-19 plays a vital role in preventing the spread of the virus. In this study, we used CXR images since X-ray imaging is cheaper, easily accessible, and faster than the conventional methods commonly used such as RT-PCR and CT. As a major contribution, the largest CXR dataset, QaTa-COV19, which consists of $2951$ COVID-19, and $116,365$ control group images, has been compiled and shared publicly as a benchmark dataset. Moreover, for the first time in the literature, we release the ground-truth segmentation masks of the infected regions along with the introduced benchmark QaTa-COV19. Furthermore, we proposed a human-machine collaborative approach, which can be used when a fast and accurate ground-truth annotation is desired but manual segmentation is slow, costly, and subjective. Finally, this study reveals the first approach ever proposed for infection map generation in CXR images. Our extensive experiments on QaTa-COV19 show that a reliable COVID-19 diagnosis can be achieved by generating infection maps, which can locate the infection on the lungs by $81.72$\% sensitivity, and $83.20$\% F1-Score. Moreover, the proposed joint approach can achieve an elegant COVID-19 detection performance with $94.96$\% sensitivity and $99.88$\% specificity. 

Many COVID-19 detectors proposed in the literature reported similar or even better detection performances. However, not only they are evaluated over small-size datasets, but also they can only discriminate between COVID-19 and normal (healthy) data, which is a straightforward task. The proposed joint approach is the only COVID-19 detector that can distinguish it from other thoracic diseases as being evaluated over the largest CXR dataset ever composed. Accordingly, the most important aspect of this study is that the generated infection maps can assist MDs for a better and objective COVID-19 assessment. For instance, it can show the time progress of the disease if the time series CXR data are generated by the proposed infection maps. It is clear that when compared with the activation maps extracted from deep models, the proposed infection maps are highly superior and reliable cues for COVID-19 infection.

\section{Data Availability}
This study introduces the QaTa-COV19 dataset, a publicly shared benchmark dataset, which is available at \href{https://www.kaggle.com/aysendegerli/qatacov19-dataset}{https://www.kaggle.com/aysendegerli/qatacov19-dataset}. The live demo of the proposed approach is implemented on \href{http://qatacov.live/}{http://qatacov.live/}.

\ifCLASSOPTIONcaptionsoff
  \newpage
\fi



%

\bibliographystyle{IEEEtran}
\bibliography{IEEEtran}

\begin{thebibliography}{10}
\providecommand{\url}[1]{#1}
\csname url@samestyle\endcsname
\providecommand{\newblock}{\relax}
\providecommand{\bibinfo}[2]{#2}
\providecommand{\BIBentrySTDinterwordspacing}{\spaceskip=0pt\relax}
\providecommand{\BIBentryALTinterwordstretchfactor}{4}
\providecommand{\BIBentryALTinterwordspacing}{\spaceskip=\fontdimen2\font plus
\BIBentryALTinterwordstretchfactor\fontdimen3\font minus
  \fontdimen4\font\relax}
\providecommand{\BIBforeignlanguage}[2]{{%
\expandafter\ifx\csname l@#1\endcsname\relax
\typeout{** WARNING: IEEEtran.bst: No hyphenation pattern has been}%
\typeout{** loaded for the language `#1'. Using the pattern for}%
\typeout{** the default language instead.}%
\else
\language=\csname l@#1\endcsname
\fi
#2}}
\providecommand{\BIBdecl}{\relax}
\BIBdecl

\bibitem{2020coronavirus}
``Severe {O}utcomes {A}mong {P}atients with {C}oronavirus {D}isease 2019
  ({COVID}-19)-{U}nited {S}tates, {F}ebruary 12-{M}arch 16, 2020. {MMWR} {M}orb
  {M}ortal {W}kly {R}ep 2020;69:343-346.'' \url{DOI:
  http://dx.doi.org/10.15585/mmwr.mm6912e2}.

\bibitem{world2020coronavirus}
{World Health Organization}, ``Coronavirus disease 2019 (covid-19): situation
  report, 88,'' 2020.

\bibitem{sohrabi2020world}
C.~Sohrabi, Z.~Alsafi, N.~O’Neill, M.~Khan, A.~Kerwan, A.~Al-Jabir,
  C.~Iosifidis, and R.~Agha, ``World health organization declares global
  emergency: A review of the 2019 novel coronavirus (covid-19),''
  \emph{International Journal of Surgery}, 2020.

\bibitem{singhal2020review}
T.~Singhal, ``A review of coronavirus disease-2019 (covid-19),'' \emph{The
  Indian Journal of Pediatrics}, pp. 1--6, 2020.

\bibitem{kakodkar2020comprehensive}
P.~Kakodkar, N.~Kaka, and M.~Baig, ``A comprehensive literature review on the
  clinical presentation, and management of the pandemic coronavirus disease
  2019 (covid-19),'' \emph{Cureus}, vol.~12, no.~4, 2020.

\bibitem{li2020stability}
Y.~Li, L.~Yao, J.~Li, L.~Chen, Y.~Song, Z.~Cai, and C.~Yang, ``Stability issues
  of rt-pcr testing of sars-cov-2 for hospitalized patients clinically
  diagnosed with covid-19,'' \emph{Journal of medical virology}, 2020.

\bibitem{tahamtan2020real}
A.~Tahamtan and A.~Ardebili, ``Real-time rt-pcr in covid-19 detection: issues
  affecting the results,'' \emph{Expert Review of Molecular Diagnostics},
  vol.~20, no.~5, pp. 453--454, 2020.

\bibitem{xia2020evaluation}
J.~Xia, J.~Tong, M.~Liu, Y.~Shen, and D.~Guo, ``Evaluation of coronavirus in
  tears and conjunctival secretions of patients with sars-cov-2 infection,''
  \emph{Journal of medical virology}, vol.~92, no.~6, pp. 589--594, 2020.

\bibitem{xiao2020false}
A.~T. Xiao, Y.~X. Tong, and S.~Zhang, ``False-negative of rt-pcr and prolonged
  nucleic acid conversion in covid-19: rather than recurrence,'' \emph{Journal
  of medical virology}, 2020.

\bibitem{yang2020laboratory}
Y.~Yang, M.~Yang, C.~Shen, F.~Wang, J.~Yuan, J.~Li, M.~Zhang, Z.~Wang, L.~Xing,
  J.~Wei \emph{et~al.}, ``Laboratory diagnosis and monitoring the viral
  shedding of 2019-ncov infections,'' \emph{MedRxiv}, 2020.

\bibitem{world2020laboratory}
{World Health Organization}, ``Laboratory testing for coronavirus disease 2019
  (covid-19) in suspected human cases: interim guidance, 2 march 2020,'' World
  Health Organization, Tech. Rep., 2020.

\bibitem{salehi2020coronavirus}
S.~Salehi, A.~Abedi, S.~Balakrishnan, and A.~Gholamrezanezhad, ``Coronavirus
  disease 2019 (covid-19): a systematic review of imaging findings in 919
  patients,'' \emph{American Journal of Roentgenology}, pp. 1--7, 2020.

\bibitem{fang2020sensitivity}
Y.~Fang, H.~Zhang, J.~Xie, M.~Lin, L.~Ying, P.~Pang, and W.~Ji, ``Sensitivity
  of chest ct for covid-19: comparison to rt-pcr,'' \emph{Radiology}, p.
  200432, 2020.

\bibitem{ai2020correlation}
T.~Ai, Z.~Yang, H.~Hou, C.~Zhan, C.~Chen, W.~Lv, Q.~Tao, Z.~Sun, and L.~Xia,
  ``Correlation of chest ct and rt-pcr testing in coronavirus disease 2019
  (covid-19) in china: a report of 1014 cases,'' \emph{Radiology}, p. 200642,
  2020.

\bibitem{bernheim2020chest}
A.~Bernheim, X.~Mei, M.~Huang, Y.~Yang, Z.~A. Fayad, N.~Zhang, K.~Diao, B.~Lin,
  X.~Zhu, K.~Li \emph{et~al.}, ``Chest ct findings in coronavirus disease-19
  (covid-19): relationship to duration of infection,'' \emph{Radiology}, p.
  200463, 2020.

\bibitem{li2020coronavirus}
Y.~Li and L.~Xia, ``Coronavirus disease 2019 (covid-19): role of chest ct in
  diagnosis and management,'' \emph{American Journal of Roentgenology}, vol.
  214, no.~6, pp. 1280--1286, 2020.

\bibitem{narin2020automatic}
A.~Narin, C.~Kaya, and Z.~Pamuk, ``Automatic detection of coronavirus disease
  (covid-19) using x-ray images and deep convolutional neural networks,''
  \emph{arXiv preprint arXiv:2003.10849}, 2020.

\bibitem{brenner2007computed}
D.~J. Brenner and E.~J. Hall, ``Computed tomography—an increasing source of
  radiation exposure,'' \emph{The New England Journal of Medicine}, vol. 357,
  no.~22, pp. 2277--2284, 2007.

\bibitem{rubin2020role}
G.~D. Rubin, C.~J. Ryerson, L.~B. Haramati, N.~Sverzellati, J.~P. Kanne,
  S.~Raoof, N.~W. Schluger, A.~Volpi, J.-J. Yim, I.~B. Martin \emph{et~al.},
  ``The role of chest imaging in patient management during the covid-19
  pandemic: a multinational consensus statement from the fleischner society,''
  \emph{Chest}, vol. 158, no.~1, pp. 106--116, 2020.

\bibitem{shi2020review}
F.~Shi, J.~Wang, J.~Shi, Z.~Wu, Q.~Wang, Z.~Tang, K.~He, Y.~Shi, and D.~Shen,
  ``Review of artificial intelligence techniques in imaging data acquisition,
  segmentation and diagnosis for covid-19,'' \emph{IEEE Reviews in Biomedical
  Engineering}, 2020.

\bibitem{chowdhury2020pdcovidnet}
N.~K. Chowdhury, M.~M. Rahman, and M.~A. Kabir, ``Pdcovidnet: a
  parallel-dilated convolutional neural network architecture for detecting
  covid-19 from chest x-ray images,'' \emph{Health information science and
  systems}, vol.~8, no.~1, pp. 1--14, 2020.

\bibitem{pham2020classification}
T.~D. Pham, ``Classification of covid-19 chest x-rays with deep learning: new
  models or fine tuning?'' \emph{Health Information Science and Systems},
  vol.~9, no.~1, pp. 1--11, 2020.

\bibitem{chowdhury2020can}
M.~E. Chowdhury, T.~Rahman, A.~Khandakar, R.~Mazhar, M.~A. Kadir, Z.~B. Mahbub,
  K.~R. Islam, M.~S. Khan, A.~Iqbal, N.~Al-Emadi \emph{et~al.}, ``Can ai help
  in screening viral and covid-19 pneumonia?'' \emph{arXiv preprint
  arXiv:2003.13145}, 2020.

\bibitem{apostolopoulos2020covid}
I.~D. Apostolopoulos and T.~A. Mpesiana, ``Covid-19: automatic detection from
  x-ray images utilizing transfer learning with convolutional neural
  networks,'' \emph{Physical and Engineering Sciences in Medicine}, p.~1, 2020.

\bibitem{hall2020finding}
L.~O. Hall, R.~Paul, D.~B. Goldgof, and G.~M. Goldgof, ``Finding covid-19 from
  chest x-rays using deep learning on a small dataset,'' \emph{arXiv preprint
  arXiv:2004.02060}, 2020.

\bibitem{wang2020covid}
L.~Wang and A.~Wong, ``Covid-net: A tailored deep convolutional neural network
  design for detection of covid-19 cases from chest x-ray images,'' \emph{arXiv
  preprint arXiv:2003.09871}, 2020.

\bibitem{sethy2020detection}
P.~K. Sethy and S.~K. Behera, ``Detection of coronavirus disease (covid-19)
  based on deep features,'' \emph{Preprints}, vol. 2020030300, p. 2020, 2020.

\bibitem{zhang2020covid}
J.~Zhang, Y.~Xie, Y.~Li, C.~Shen, and Y.~Xia, ``Covid-19 screening on chest
  x-ray images using deep learning based anomaly detection,'' \emph{arXiv
  preprint arXiv:2003.12338}, 2020.

\bibitem{afshar2020covid}
P.~Afshar, S.~Heidarian, F.~Naderkhani, A.~Oikonomou, K.~N. Plataniotis, and
  A.~Mohammadi, ``Covid-caps: A capsule network-based framework for
  identification of covid-19 cases from x-ray images,'' \emph{arXiv preprint
  arXiv:2004.02696}, 2020.

\bibitem{yamac2020convolutional}
M.~Yamac, M.~Ahishali, A.~Degerli, S.~Kiranyaz, M.~E. Chowdhury, and
  M.~Gabbouj, ``Convolutional sparse support estimator based covid-19
  recognition from x-ray images,'' \emph{arXiv preprint arXiv:2005.04014},
  2020.

\bibitem{ahishali2020comparative}
M.~Ahishali, A.~Degerli, M.~Yamac, S.~Kiranyaz, M.~E. Chowdhury, K.~Hameed,
  T.~Hamid, R.~Mazhar, and M.~Gabbouj, ``Advance warning methodologies for
  covid-19 using chest x-ray images,'' \emph{arXiv preprint arXiv:2006.05332},
  2020.

\bibitem{shi2020large}
F.~Shi, L.~Xia, F.~Shan, D.~Wu, Y.~Wei, H.~Yuan, H.~Jiang, Y.~Gao, H.~Sui, and
  D.~Shen, ``Large-scale screening of covid-19 from community acquired
  pneumonia using infection size-aware classification,'' \emph{arXiv preprint
  arXiv:2003.09860}, 2020.

\bibitem{yeh2020cascaded}
C.-F. Yeh, H.-T. Cheng, A.~Wei, K.-C. Liu, M.-C. Ko, P.-C. Kuo, R.-J. Chen,
  P.-C. Lee, J.-H. Chuang, C.-M. Chen \emph{et~al.}, ``A cascaded learning
  strategy for robust covid-19 pneumonia chest x-ray screening,'' \emph{arXiv
  preprint arXiv:2004.12786}, 2020.

\bibitem{oh2020deep}
Y.~Oh, S.~Park, and J.~C. Ye, ``Deep learning covid-19 features on cxr using
  limited training data sets,'' \emph{IEEE Transactions on Medical Imaging},
  2020.

\bibitem{ozturk2020automated}
T.~Ozturk, M.~Talo, E.~A. Yildirim, U.~B. Baloglu, O.~Yildirim, and U.~R.
  Acharya, ``Automated detection of covid-19 cases using deep neural networks
  with x-ray images,'' \emph{Computers in Biology and Medicine}, p. 103792,
  2020.

\bibitem{alom2020covid_mtnet}
M.~Z. Alom, M.~Rahman, M.~S. Nasrin, T.~M. Taha, and V.~K. Asari,
  ``Covid\_mtnet: Covid-19 detection with multi-task deep learning
  approaches,'' \emph{arXiv preprint arXiv:2004.03747}, 2020.

\bibitem{haghanifar2020covid}
A.~Haghanifar, M.~M. Majdabadi, and S.~Ko, ``Covid-cxnet: Detecting covid-19 in
  frontal chest x-ray images using deep learning,'' \emph{arXiv preprint
  arXiv:2006.13807}, 2020.

\bibitem{shan2020lung}
F.~Shan, Y.~Gao, J.~Wang, W.~Shi, N.~Shi, M.~Han, Z.~Xue, and Y.~Shi, ``Lung
  infection quantification of covid-19 in ct images with deep learning,''
  \emph{arXiv preprint arXiv:2003.04655}, 2020.

\bibitem{zhang2020clinically}
K.~Zhang, X.~Liu, J.~Shen, Z.~Li, Y.~Sang, X.~Wu, Y.~Zha, W.~Liang, C.~Wang,
  K.~Wang \emph{et~al.}, ``Clinically applicable ai system for accurate
  diagnosis, quantitative measurements, and prognosis of covid-19 pneumonia
  using computed tomography,'' \emph{Cell}, 2020.

\bibitem{qiu2020miniseg}
Y.~Qiu, Y.~Liu, and J.~Xu, ``Miniseg: An extremely minimum network for
  efficient covid-19 segmentation,'' \emph{arXiv preprint arXiv:2004.09750},
  2020.

\bibitem{vaya2020bimcv}
M.~d. l.~I. Vay{\'a}, J.~M. Saborit, J.~A. Montell, A.~Pertusa, A.~Bustos,
  M.~Cazorla, J.~Galant, X.~Barber, D.~Orozco-Beltr{\'a}n, F.~Garcia
  \emph{et~al.}, ``Bimcv covid-19+: a large annotated dataset of rx and ct
  images from covid-19 patients,'' \emph{arXiv preprint arXiv:2006.01174},
  2020.

\bibitem{covidimage}
\BIBentryALTinterwordspacing
``{COVID}-19 {I}mage {R}epository,'' 2020. [Online]. Available:
  \url{https://github.com/ml-workgroup/covid-19-image-repository. [Accessed on
  16-September-2020]}
\BIBentrySTDinterwordspacing

\bibitem{coviddatabase}
\BIBentryALTinterwordspacing
``{COVID-19 DATABASE},'' 2020. [Online]. Available:
  \url{https://www.sirm.org/category/senza-categoria/covid-19/. [Accessed on
  16-September-2020]}
\BIBentrySTDinterwordspacing

\bibitem{cohen2020covid}
J.~P. Cohen, P.~Morrison, and L.~Dao, ``Covid-19 image data collection,''
  \emph{arXiv preprint arXiv:2003.11597}, 2020.

\bibitem{radiodatabase}
\BIBentryALTinterwordspacing
``{COVID-19} {R}adiography {D}atabase,'' 2020. [Online]. Available:
  \url{https://www.kaggle.com/tawsifurrahman/covid19-radiography-database.
  [Accessed on 16-September-2020]}
\BIBentrySTDinterwordspacing

\bibitem{chestimaging}
\BIBentryALTinterwordspacing
``{C}hest {I}maging,'' 2020. [Online]. Available:
  \url{https://www.eurorad.org/. [Accessed on 16-September-2020]}
\BIBentrySTDinterwordspacing

\bibitem{RSNA}
\BIBentryALTinterwordspacing
``{RSNA} {P}neumonia {D}etection {C}hallenge,'' 2018. [Online]. Available:
  \url{https://www.kaggle.com/c/rsna-pneumonia-detection-challenge/overview.
  [Accessed on 22-September-2020]}
\BIBentrySTDinterwordspacing

\bibitem{bustos2020padchest}
A.~Bustos, A.~Pertusa, J.-M. Salinas, and M.~de~la Iglesia-Vay{\'a},
  ``Padchest: A large chest x-ray image dataset with multi-label annotated
  reports,'' \emph{Medical Image Analysis}, p. 101797, 2020.

\bibitem{kermany2018identifying}
D.~S. Kermany, M.~Goldbaum, W.~Cai, C.~C. Valentim, H.~Liang, S.~L. Baxter,
  A.~McKeown, G.~Yang, X.~Wu, F.~Yan \emph{et~al.}, ``Identifying medical
  diagnoses and treatable diseases by image-based deep learning,'' \emph{Cell},
  vol. 172, no.~5, pp. 1122--1131, 2018.

\bibitem{indiana}
D.~Demner-Fushman, M.~D. Kohli, M.~B. Rosenman, S.~E. Shooshan, L.~Rodriguez,
  S.~Antani, G.~R. Thoma, and C.~J. McDonald, ``Preparing a collection of
  radiology examinations for distribution and retrieval,'' \emph{Journal of the
  American Medical Informatics Association}, vol.~23, no.~2, pp. 304--310,
  2016.

\bibitem{chinaUS}
S.~Jaeger, S.~Candemir, S.~Antani, Y.-X.~J. W{\'a}ng, P.-X. Lu, and G.~Thoma,
  ``Two public chest x-ray datasets for computer-aided screening of pulmonary
  diseases,'' \emph{Quantitative imaging in medicine and surgery}, vol.~4,
  no.~6, p. 475, 2014.

\bibitem{wang2017chestx}
X.~Wang, Y.~Peng, L.~Lu, Z.~Lu, M.~Bagheri, and R.~M. Summers, ``Chestx-ray8:
  Hospital-scale chest x-ray database and benchmarks on weakly-supervised
  classification and localization of common thorax diseases,'' in
  \emph{Proceedings of the IEEE conference on computer vision and pattern
  recognition}, 2017, pp. 2097--2106.

\bibitem{ronneberger2015u}
O.~Ronneberger, P.~Fischer, and T.~Brox, ``U-net: Convolutional networks for
  biomedical image segmentation,'' in \emph{International Conference on Medical
  image computing and computer-assisted intervention}.\hskip 1em plus 0.5em
  minus 0.4em\relax Springer, 2015, pp. 234--241.

\bibitem{zhou2018unet++}
Z.~Zhou, M.~M.~R. Siddiquee, N.~Tajbakhsh, and J.~Liang, ``Unet++: A nested
  u-net architecture for medical image segmentation,'' in \emph{Deep Learning
  in Medical Image Analysis and Multimodal Learning for Clinical Decision
  Support}.\hskip 1em plus 0.5em minus 0.4em\relax Springer, 2018, pp. 3--11.

\bibitem{yu2018deep}
F.~Yu, D.~Wang, E.~Shelhamer, and T.~Darrell, ``Deep layer aggregation,'' in
  \emph{Proceedings of the IEEE conference on computer vision and pattern
  recognition}, 2018, pp. 2403--2412.

\bibitem{huang2017densely}
G.~Huang, Z.~Liu, L.~Van Der~Maaten, and K.~Q. Weinberger, ``Densely connected
  convolutional networks,'' in \emph{Proceedings of the IEEE conference on
  computer vision and pattern recognition}, 2017, pp. 4700--4708.

\bibitem{rajpurkar2017chexnet}
P.~Rajpurkar, J.~Irvin, K.~Zhu, B.~Yang, H.~Mehta, T.~Duan, D.~Ding, A.~Bagul,
  C.~Langlotz, K.~Shpanskaya \emph{et~al.}, ``Chexnet: Radiologist-level
  pneumonia detection on chest x-rays with deep learning,'' \emph{arXiv
  preprint arXiv:1711.05225}, 2017.

\bibitem{szegedy2016rethinking}
C.~Szegedy, V.~Vanhoucke, S.~Ioffe, J.~Shlens, and Z.~Wojna, ``Rethinking the
  inception architecture for computer vision,'' in \emph{Proceedings of the
  IEEE conference on computer vision and pattern recognition}, 2016, pp.
  2818--2826.

\bibitem{he2016deep}
K.~He, X.~Zhang, S.~Ren, and J.~Sun, ``Deep residual learning for image
  recognition,'' in \emph{Proceedings of the IEEE conference on computer vision
  and pattern recognition}, 2016, pp. 770--778.

\bibitem{lin2017focal}
T.-Y. Lin, P.~Goyal, R.~Girshick, K.~He, and P.~Doll{\'a}r, ``Focal loss for
  dense object detection,'' in \emph{Proceedings of the IEEE international
  conference on computer vision}, 2017, pp. 2980--2988.

\bibitem{milletari2016v}
F.~Milletari, N.~Navab, and S.-A. Ahmadi, ``V-net: Fully convolutional neural
  networks for volumetric medical image segmentation,'' in \emph{2016 fourth
  international conference on 3D vision (3DV)}.\hskip 1em plus 0.5em minus
  0.4em\relax IEEE, 2016, pp. 565--571.

\bibitem{abadi2016tensorflow}
M.~Abadi, A.~Agarwal, P.~Barham, E.~Brevdo, Z.~Chen, C.~Citro, G.~S. Corrado,
  A.~Davis, J.~Dean, M.~Devin \emph{et~al.}, ``Tensorflow: Large-scale machine
  learning on heterogeneous distributed systems,'' \emph{arXiv preprint
  arXiv:1603.04467}, 2016.

\bibitem{kingma2014adam}
D.~P. Kingma and J.~Ba, ``Adam: A method for stochastic optimization,''
  \emph{arXiv preprint arXiv:1412.6980}, 2014.

\bibitem{selvaraju2017grad}
R.~R. Selvaraju, M.~Cogswell, A.~Das, R.~Vedantam, D.~Parikh, and D.~Batra,
  ``Grad-cam: Visual explanations from deep networks via gradient-based
  localization,'' in \emph{Proceedings of the IEEE international conference on
  computer vision}, 2017, pp. 618--626.

\end{thebibliography}




\end{document}